\documentclass[preprint,superscriptaddress,showpacs,amsmath,amssymb,aps,prx,floatfix]{revtex4-1}
\usepackage[utf8]{inputenc}
\usepackage{amsmath}
\usepackage{xcolor}
\usepackage{amsfonts}
\usepackage{amssymb}
\usepackage{graphicx}
\usepackage[T1]{fontenc}
\usepackage{csvsimple}
\usepackage{hyperref}% add hypertext capabilities
\usepackage{nameref}
\usepackage{longtable}
\usepackage{booktabs}
\usepackage{soul}
\usepackage{xr}
\usepackage{setspace}
\makeatletter
\usepackage[margin=1in]{geometry}
\newcommand*{\addFileDependency}[1]{
  \typeout{(#1)}
  \@addtofilelist{#1}
  \IfFileExists{#1}{}{\typeout{No file #1.}}
}
\makeatother

\newcommand*{\myexternaldocument}[1]{
    \externaldocument{#1}
    \addFileDependency{#1.tex}
    \addFileDependency{#1.aux}
}
\myexternaldocument{sm}

\begin{document}
	
\title{The impact of inter-city mobility on urban welfare}

\newcommand{\RochesterP}{Department of Physics \& Astronomy, University of Rochester, Rochester, NY 14607, USA}
\newcommand{\ZaragozaP}{GOTHAM Lab‚ Department of Condensed Matter Physics and Institute for Biocomputation and Physics of Complex Systems (BIFI),	University of Zaragoza, E-50009 Zaragoza, Spain.}
\newcommand{\MichiganP}{Department   of   Physics,   University   of   Michigan,   Ann   Arbor,  MI 48109, USA}
\newcommand{\ExeterP}{Department of Computer Science, University of Exeter, Exeter, UK}
	
\newcommand{\rev}[1]{{\color{red} #1}}

\author{Sayat Mimar}                    
\affiliation{\RochesterP}
\author{David Soriano-Pa\~nos}                    
\affiliation{\ZaragozaP}
\author{Alec Kirkley}                    
\affiliation{\MichiganP}
\affiliation{School of Data Science, City University of Hong Kong, Hong Kong}
\author{Hugo Barbosa}
\affiliation{\ExeterP}
\author{Adam Sadilek}
\thanks{sadilekadam@google.com}
\affiliation{Google Inc., 1600 Amphitheatre Parkway, Mountain View, CA, 94043, USA}
\author{Alex Arenas}   
\thanks{alexandre.arenas@urv.cat}
\affiliation{Department d'Enginyeria Inform\'atica i Matem\`atiques, Universitat Rovira i Virgili, 43007 Tarragona (Spain).}
\author{J. G\'omez-Garde\~nes}
\thanks{gardenes@unizar.es}
\affiliation{\ZaragozaP}
\author{Gourab Ghoshal}     
\thanks{gghoshal@pas.rochester.edu}
\affiliation{\RochesterP}

\begin{abstract}
\singlespacing
While much effort has been devoted to understand the role of intra-urban characteristics on sustainability and growth, much remains to be understood about the effect of inter-urban interactions and the role cities have in determining each other's urban welfare. Here we consider a global mobility network of population flows between cities as a proxy for the communication between these regions, and analyze how these flows impact socioeconomic indicators that measure economic success. We use several measures of centrality to rank cities according to their importance in the mobility network, finding PageRank to be the most effective measure for reflecting these prosperity indicators. Our analysis reveals that the characterization of the welfare of cities based on mobility information hinges on their corresponding development stage. Namely, 
while network-based predictions of welfare correlate well with economic indicators in mature cities, for developing urban areas additional information about the prosperity of their mobility neighborhood is needed. For these developing cities, those that are connected to sets of mature cities show markedly better socio-economic indicators than those connected to less mature cities. We develop a simple generative model for the allocation of population flows out of a city that balances the costs and benefits of interaction with other cities that are successful, finding that it provides a strong fit to the flows observed in the global mobility network and highlights the differences in flow patterns between developed and developing urban regions. Our results hint towards the importance of leveraging inter-urban connections in service of urban development and welfare.
\end{abstract}
\maketitle
%the cities that these urban areas interact with in the mobility network. Our analysis reveals at high granularity that mobility network-based predictions of success alone correlate well with economic indicators for mature cities that form the backbone of the mobility network. Meanwhile, to make good predictions about the economic success of developing urban areas that are situated in the periphery of the network, we require 
%We also find that it is critical to account for both the population flows between cities and the corresponding distances of these flows as a means of capturing the demand (or cost) for interaction between urban areas, as this significantly strengthens the association between centrality and city prosperity. 
%We develop a simple generative model for the allocation of population flows out of a city that balances the costs and benefits of interaction with other cities that are successful, finding that it provides a strong fit to the flows observed in the global mobility network and highlights the differences in flow patterns between developed and developing urban regions.	
\section{Introduction}
Given the recent trend of rapid urbanization wherein the majority of the global population now resides in urban centers~\cite{DESAUN2018, urbanization}, cities are at the center of innovation and technological advancement~\cite{berry2008urbanization,Le2012}. The current relevance of cities has fueled the birth of the so-called \emph{science of cities} aimed at uncovering the physical and structural features driving their growth and function~\cite{Batty2013, Pan2013, Barthelemy2016}. Among the different topics addressed by this discipline, one recurrent question has been that of quantifying how human mobility shapes urban dynamics and provides reliable information on different socioeconomic indicators. Along this line, recent studies have used intra-city mobility flows to understand spatial city organization and its connection to urban characteristics such as facilities, jobs and services that are crucial for city livability and sustainability \cite{Bassolas2019,Lee2017}. Other works have investigated accessibility in urban systems considering city topology and infrastructure that mediate interactions and activities of inhabitants \cite{Shelton2015, Xu2018,Barbosa_2021}. More generally, mobility flows have been used to study the structure of urban areas and dynamics taking place within them  \cite{Louail2015,Roth2011,Barbosa2018,Kirkley2018, verbavatz2020growth}. 

The welfare of a city is just one example of the emergence of success within a system composed of a large set of similar elements among which there are relations of competition and cooperation. The study of the mechanisms driving the emergence of cases of success has been recently addressed in a variety of fields. For example, luck and randomness have been found to be crucial features behind the prosperity of scientific~\cite{Wang2013,Sinatra2016} or creative careers~\cite{Janosov2020}. Apart from the impact of intrinsically stochastic events, the roots of individual success in different disciplines has been addressed through the lens of network science. Specifically, in \cite{Fraiberger2018}, the authors capture institutional prestige by studying a co-exhibition network where nodes represent art galleries and the links are determined by the movement of artists among institutions. Similarly, in a recent work \cite{Bonaventura2020}, the network of professional relationships of start-ups is built with the time-varying flows of employees among world-wide companies. This network-based approach has shown to have predictive power of the future success of companies. Finally, in \cite{Bonaventura2019}, the authors construct a Workforce Mobility Network among metropolitan areas in the US and predict diverse socioeconomic outputs of urban areas such as number of New Patents, the number of R$\&$D establishments, and Total Wages.

``What makes a city successful?" is an intriguing and complicated question as there are multiple factors shaping the success of urban systems. Here we aim at addressing this question by leveraging the close relationship between human mobility and the functioning of cities. To this end, we propose a network science approach to shed light on the heterogeneous distribution of success, as measured by various socioeconomic indicators, observed among cities worldwide. We use anonymous and aggregated inter-city flows to construct a global mobility network between cities, with the aim of capturing this success using the observed flow patterns. We start by ranking cities with respect to various network centrality measures---in particular connectivity, PageRank, and eigenvector centrality---which capture not only the strength of connections but also their quality, and compare these centralities with socioeconomic indicators in $268$ global metropolitan areas. We demonstrate that the welfare of an urban area is largely influenced by the welfare of its neighborhood, that is, those cities with which it has the highest volume of inter-urban population flows. This impact becomes even more visible when grouping cities into various categories according to geographic region or development level, to distinguish flow profiles of cities in different groups. We observe that the international connections for a city in a less developed category (or geographic region) contribute significantly to its prosperity. Finally, we build a simple model for the outgoing flows from a city, finding that the success of a city is a major element in attracting flows from others, and that less developed cities place greater priority on directing flows to highly successful cities.

\section{Results}
\subsection{Data Description}
\subsubsection{Mobility Data}
The Google Aggregated Mobility Research Dataset contains anonymized mobility flows aggregated over users who have turned on the Location History setting, which is off by default. This is similar to the data used to show how busy certain types of places are in Google Maps, which helps to identify when a local business tends to be the most crowded. The dataset aggregates flows of people from region to region weekly.

To produce this dataset, machine learning methods are applied to logs data to automatically segment it into semantic trips \footnote{\url{https://www.nature.com/articles/s41467-019-12809-y}}. To provide strong privacy guarantees, all trips are anonymized and aggregated using a differentially private mechanism \footnote{\url{https://research.google/pubs/pub48778/}} to aggregate flows over time  \footnote{\url{https://policies.google.com/technologies/anonymization}}. The analysis in this paper is done on the resulting heavily aggregated and differentially private data. No individual user data was ever manually inspected, only heavily aggregated flows of large populations were handled.

All anonymized trips are processed in aggregate to extract their origin and destination location and time. For example, if $n$ unique users traveled from location $a$ to location $b$ within week $w$, the corresponding cell $(a,b,w)$ in the mobility tensor would have a value of $n \pm \eta$ where $\eta$ is noise drawn from a Laplace distribution with mean $0$ and scale $1/0.66$. We then remove all metrics for which the noisy number of users is lower than $100$, following the process described in \footnote{\url{https://research.google/pubs/pub48778/}}. This automated Laplace mechanism yields a ($\epsilon$, $\delta$)-differential privacy guarantee of $\epsilon$ = 0.66 and $\delta$ =$2.1\times10^{-29}$ per metric. The parameter $\epsilon$ controls the noise intensity in terms of its variance, while $\delta$ represents the deviation from pure $\epsilon$-privacy. The closer each value is to zero, the stronger the privacy guarantees. The resulting mobility flows are encoded in an origin-destination matrix ${\bf T}$ whose elements $T_{ij}$ represent the flows from location $i$ to $j$  aggregated over the full year.

\subsubsection{Socioeconomic indicators}
To quantify the economic success of cities, we collected data of city-level socioeconomic variables from the company Jones Lang LaSalle IP, Inc.~(JLL) for the year 2018 \cite{jll-data}. The report published by JLL gives the GDP, Total Real Estate Investment, and a Cross-Border Real Estate Investment and Commercial Attraction Index (BHI) for $300$ cities worldwide. In particular, the BHI is a composite measure that accounts for key real estate indicators (investment volumes and commercial real estate stock), as well as socio-economic and business indicators such as economic output, population, corporate presence, and air connectivity.
\par

Besides these continuous variables, there are categorical variables in the dataset that divide cities into different subsets based on their level of development and geographical region. The indicated levels of development (in increasing order) are \textit{Early Growth}, \textit{Developing}, \textit{Transitional}, and \textit{Mature}, which are assigned following information concerning the city's real estate liquidity, the depth of its corporate occupier base, and the quality and range of its commercial stock. The geographical sub-regions indicated are \textit{North America}, \textit{LATAM $\&$ Caribbean}, \textit{Asia},
\textit{Australasia}, \textit{Western Europe}, \textit{CEE/CIS}, \textit{MENA}, and
\textit{Sub-Saharan Africa}. For our analysis we consider 268 metropolitan areas, excluding China for which we do not have mobility data. (See Figs. S1, S2 and Table S1 for more details about the urban areas used in this study.)

\subsection{Relating welfare and PageRank centrality}

\begin{figure}[t!]	

	\centering
	\includegraphics[width=1.0\linewidth]{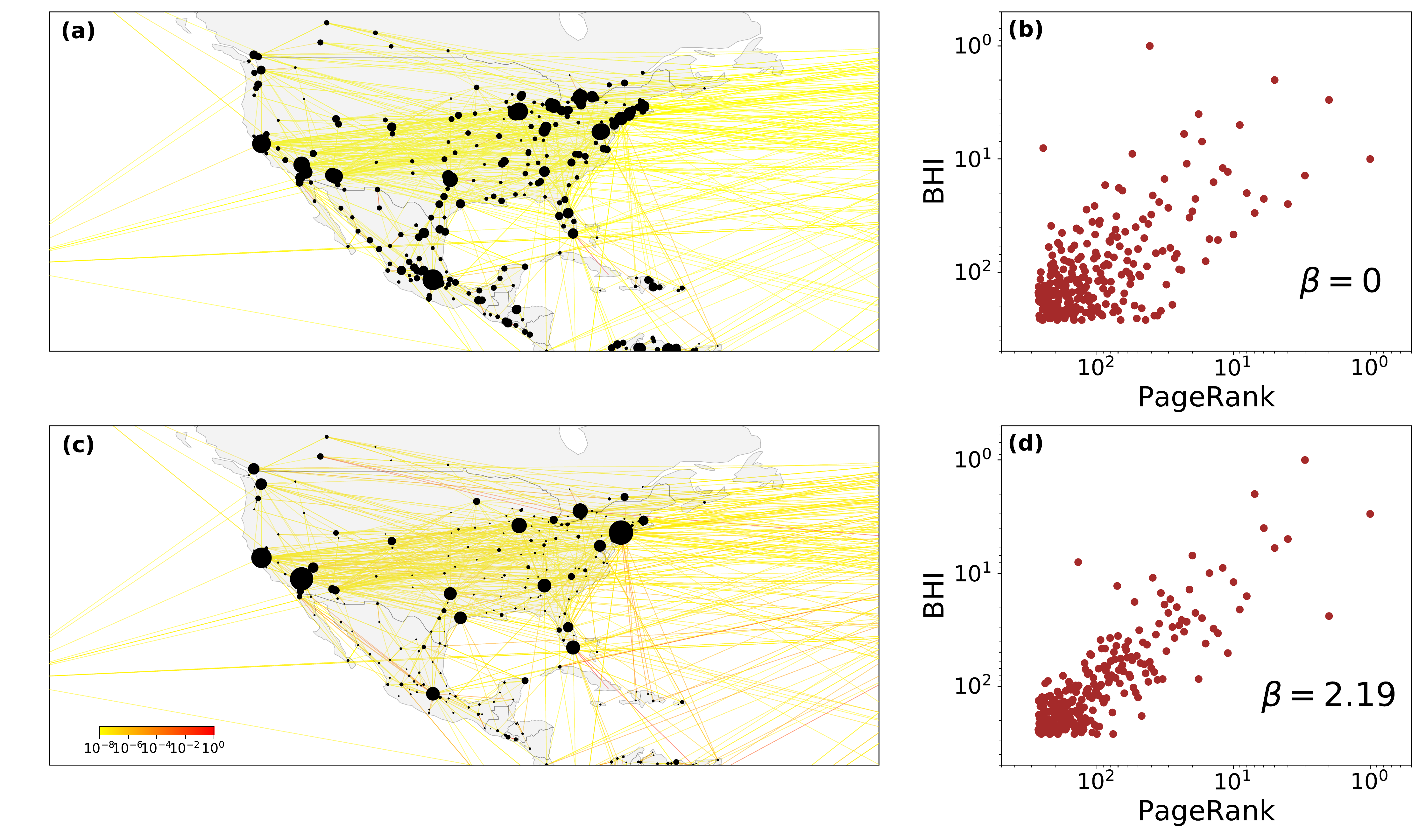}\\
	\caption{{\bf Connecting network position and welfare}. Weights $W_{ij}$, of the global mobility network for \textbf{a} the raw mobility flows ($\beta = 0$ in Eq.~\eqref{eq:weights}) and \textbf{c} when incorporating the effect of distance in the weights ($\beta = 2.19$), shown for North American cities. Node sizes are proportional to their PageRank in both maps. By including distance in the flows, we give stronger weight to long-range trips. \textbf{b} BHI vs PageRank for $\beta =0$, with a Spearman correlation coefficient $\rho_s  = 0.53$. \textbf{d} BHI vs PageRank for $\beta = 2.19$, yielding $\rho_s  = 0.77$.}
\label{fig:map}	
\end{figure}

While we have metadata for 268 cities, for the purposes of our analysis, we analyze the mobility network of $N=1,774$ nodes representing global metropolitan areas with population $P > 100,000$ and $E=39,868$ directed edges. To incorporate, in a tunable way, the contribution of long-distance travel to economic activity, we define the edge weights of the mobility network as:
\begin{equation}
W_{ij} = T_{ij} \times d_{ij}^{\beta},
\label{eq:weights}
\end{equation}
where $\beta\geq 0$. In this expression $T_{ij}$ is the total flow from city $i$ to $j$ and the $d_{ij}$ is the Euclidean distance between them, which allows us to account for the effect of distance as a cost in population movement \cite{Simini2012,Barthelemy2011}. Incorporating this allows for including the fact  that long-range connections indicate greater levels of implied connectivity between the cities, given the associated cost of travel. The optimal value of the exponent $\beta$ is determined later. The effect of distance on the flow-weights is illustrated in Fig.~\ref{fig:map} where panel \textbf{a} shows mobility flows only ($\beta = 0$), whereas in panel \textbf{b} we show an illustrative example for $\beta > 0$. As the figure shows, incorporating the effect of distance makes long-range flows more prominent; colors indicate magnitude of flow, increasing from yellow to red, and node sizes correspond to PageRank. (Maps generated using Shapely \footnote{\url{https:// pypi. org/ proje ct/ Shape ly/}} and GeoPandas \footnote{\url{https:// geopa ndas. org/}} packages in Python.)

\par
Next we rank cities according to their importance based on their position in the mobility network. An ideal metric for that is the PageRank centrality, a network-based diffusion algorithm, used by Google and other web search engines to estimate the importance or quality 
of web pages. The algorithm takes into account both the number of connections as well as the importance of the neighbors a node connects to~\cite{Page1998, Ghoshal2011}. For a weighted directed network, the PageRank of a node $i$ is computed as~\cite{10.5555/998669.998911}:
\begin{equation}
PR_{i}=\frac{(1-\alpha)}{N}+\alpha \sum_{j} \frac{{W}_{ji}PR_{j}  }{ s^{\mathrm{out}}_j},
\label{eq:PageRank}
\end{equation}
where $s^{\mathrm{out}}_j$ is the total strength of outgoing links from node $j$ and $\alpha$ is a ``reset" parameter in the range $[0,1]$ (we use $\alpha = 0.85$). In addition to webpages, the metric has been successfully employed to rank scientists based on their citation patterns~\cite{Walker2007,Chen2007}, disease-causing genes based on protein-protein interactions~\cite{Chen_2009}, roads or streets in terms of traffic~\cite{Jiang_2008}, ecological species based on their position in the food web~\cite{Allesina2009}, highlight cancer genes in proteomic data~\cite{Gabor_2011} among other applications.

\par
%\begin{figure}[t!]	

%	\centering
%	\includegraphics[width=1.0\linewidth]{conectivity-PageRank.pdf}\\
%	\caption{\textbf{a} Weighted connectivity vs BHI with spearman correlation coefficient $\rho_s = 0.66$.\textbf{b} PageRank vs BHI with Spearman's correlation coefficient $\rho_s = 0.69$}
%\label{fig:conpage}	
%\end{figure}

We start our analysis by ranking cities in the mobility network by their value of PageRank and plot it against one of the composite socioeconomic indicators, the BHI. The strength of association is measured by the Spearman's correlation  coefficient  $\rho_s$ that  measures  the  strength  and  direction  of  association  between two ranked variables. In Fig.~\ref{fig:map}{\bf c} we plot the case for $\beta = 0$ finding a monotonic relationship with $\rho_s = 0.53$. Including distance in the edge-weights reveals an increase in the association between the two variables for any non-zero $\beta$. In Fig. S3 panels \textbf{a} and \textbf{b} we plot $\rho_s$ as well as the Pearson's correlation coefficient $\rho_p$ between $\log(\text{BHI})$ and $\log(PR)$ as a function of $\beta$, finding a peak value for both correlation coefficients at $\beta=2.19$ ($\rho_s$) and $\beta=1.88$ ($\rho_p$). 
%Hence, PageRank has higher predictive power than strength as it is known that there is a set of super-stable nodes with high connectivity such that their ranking is independent of their neighbors \cite{Ghoshal2011}. However, for nodes with lower degree, the quality of connections become more important. 
In Fig.~\ref{fig:map} panel \textbf{d} we plot the scatter plot for PageRank and BHI for $\beta = 2.19$, indicating a much stronger correlation ($\rho_s = 0.77$). In Fig. S4 we show the connection of PageRank with other socioeconomic indicators (total real estate investment, GDP, and cross-border real estate investment) for a variety of choices of the edge-weights: the raw flow $T_{ij}$, the euclidean distance $d_{ij}$ as well as Eq.~\eqref{eq:weights}. The results indicate that (i) PageRank is correlated with all socioeconomic measures (the strongest being BHI) and (ii) this association is enhanced when both the flow and distance is incorporated in the edge-weights. We also show the results for two other centrality measures (weighted degree and eigenvector centrality).  Taken together, it appears that both the position of a city in the global mobility network, as well as the strength and number of connections is a reasonable predictor of its welfare.
\par

\begin{figure}[t!]	
	\centering
	\includegraphics[width=0.82\linewidth]{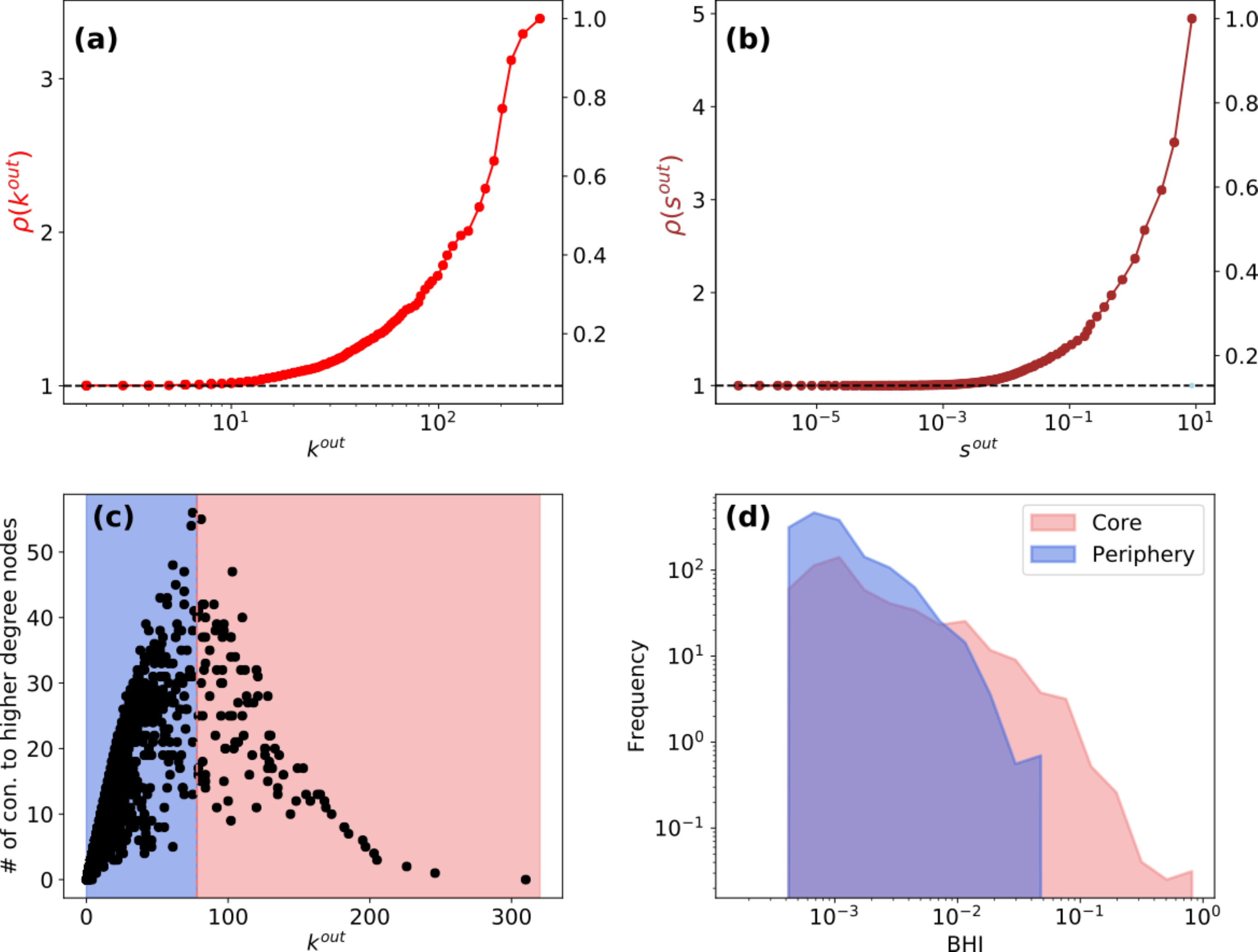}
	\caption{{\bf Rich-club organization and core-periphery structure of the mobility network}. The richness parameter $\rho(r)$ (Eq.~\eqref{eq:richness_ratio})  \textbf{a} for the connectivity $r=k^{out}$ and \textbf{b} weighted out-degree $r=s^{\text{out}}$ . The monotonically  increasing trend of $\rho(r)$ with $r$ indicates subsets of high-connectivity cities that connect to each other more than would be expected as merely a consequence of the distribution of links and edge-weights. \textbf{c} The number of neighbors of each node, with a higher out-degree than  the node itself, as a function of the node's out-degree. The scatter plot has a peak separating cities into two clusters of core (in red) and peripheral (in blue) nodes. \textbf{d} The distribution of BHI in the core and peripheral cities, showing that cities in the core typically have much higher values of BHI than those in the periphery.}
	\label{fig:richness}
\end{figure}

%(see \ref{sifig:summary} for correlation of other centrality measures with socioeconomic indicators).
Next, we dis-aggregate the cities based on their level of development and geographical region to check whether the results are consistent across these categories. In Fig. S5 we plot the PageRank against BHI for four levels of development: Mature, Transitional, Developing and Early Growth. We find a progressive decrease in the strength of association from Mature cities ($\rho_s = 0.86$) through transitional and developing cities ($\rho_s = 0.73, 0.88$, respectively) and finally a weak correlation in early growth cities ($\rho_s = 0.38$). Indeed, the differences become rather stark when looking at the other socioeconomic indicators as seen in Fig. S7. In mature cities the PageRank is strongly correlated with all indicators, in transitional cities the correlation with real-estate investment weakens significantly, and moving to the other two categories the correlation is weak with all indicators.
Differences also exist when grouping cities based on their geographical location as shown in Fig. S6. In prosperous regions (North America, Australasia and Western Europe) the association between PageRank and BHI is quite strong ($\rho_s = 0.86, 0.93,  0.76$ respectively), in areas with mixed levels of prosperity such as Latin America, Asia and the Middle East, the correlations are slightly weaker ($\rho_s = 0.72, 0.86, 0.73$) with the weakest correlations being in Sub-Saharan Africa ($\rho_s = 0.38$). 

\subsection{Effect of position in the mobility network}

To check whether these observed differences based on maturity and location are consequences of higher-order topological features of the mobility network, we next  employ the \textit{k-core} decomposition, that identifies strongly connected core-nodes and sparsely connected peripheral nodes, revealing meso-scale network structure~\cite{Zhang2015,Rombach2014}. Applying this analysis we find two core clusters consisting of 49 Western European cities and 40 North American cities as shown in Fig. S8. The rest of the clusters are much smaller (sizes between 5-20) and lie in the peripheral layers of the mobility network.  The role of a city's position in the inner or outer-layers becomes clear when plotting $\rho_s$ between PageRank and BHI as a function of the layer-number $k$ (numbered in increasing order from outer- to inner-layer) as shown in Fig. S9. We see a monotonically decreasing trend of the strength of correlation as one moves from the core to the peripheral layers indicating that PageRank become a poor predictor of welfare for cities located more in the outer-layers of the mobility network.
\par
The existence of large core clusters concentrated in Western Europe and North America (all Mature or Transitional cities) suggest that they form a densely interconnected backbone of the mobility network and thus control most of human capital and resource flow. This can be quantified by studying the weighted rich-club organization~\cite{Opsahl2008} which is defined as
\begin{equation}
\rho(r)=\frac{\phi(r)}{\phi_{\text {null }}(r)},
\label{eq:richness_ratio}
\end{equation}
where $r$ is a \emph{richness parameter}. Two flavors of the parameter that we consider are the out-degree $k^{\text{out}}$ and the out-strength $s^{\text{out}}$. Given this, $\phi(k)$ is a fraction that measures the number of edges between nodes of degree $\geq k^{\text{out}}$ compared to the edges they would share if all these nodes were connected to each other. The quantity $\phi_{\text {null }}(k^{\text{out}})$ is the corresponding measure under out degree-preserving reshuffling of links. For $s^{\text{out}}$, $\phi_{\text{null}}(s^{\text{out}})$ is computed through both edge and weight reshuffling. Values of $\rho(r) > 1$ indicate a set of nodes that are connected to more cities, with higher exchange of populations, than one would expect merely as a consequence of the distribution of links and edge-weights. In Fig.~\ref{fig:richness}\textbf{a}, we plot $\rho(k^{\text{out}})$ and in \textbf{b} $\rho(s^{\text{out}})$, finding that nodes with high connectivity and strength of connections are between two to three times more connected to each other than would be expected by random chance. The results indicate that Mature and Transitional cities form a \emph{rich-club} dominating mobility flows in the network.  

Finally, while the $k$-core decomposition allows for assigning nodes to layers, there is no clear way to define a boundary between the core and periphery. To do so, we use a method introduced in \cite{Ma2015}: First, we sort nodes with respect to $k^{\text{out}}$ in increasing order. Then, for each node, we count the number of its neighbors that have out-degree higher than itself. In Fig. ~\ref{fig:richness}\textbf{c}, we plot the sorted degrees of all nodes and their corresponding number of connections to higher-degree nodes. There is a clear peak in the scatter plot that allows us to distinguish 2 clusters of cities being in the core (in red) and in the periphery (in blue) \cite{Guo2019}. The figure indicates that the top $~30\%$ of nodes in terms of their out-degree are in the core of the mobility network, consistent with the two dense cores identified by the $k-$core method in Fig. S9. Having identified the core and periphery we plot the distribution of BHI in the two regions of the network in Fig.~\ref{fig:richness}{\bf d} finding that cities in the core are in general more prosperous than peripheral cities.

\subsection{Interplay between welfare and mobility neighborhood}

The results thus far point towards strongly inter-connected sets of core cities that dominate  mobility flow (and therefore exchange of human resources) that are successful in terms of socioeconomic indicators. This success can be reasonably well predicted by their centrality, or just their topological features in relation to the mobility network. For cities in the periphery, and those that are not so tightly connected, their level of development cannot be simply captured by topological aspects. In particular PageRank is a poor measure to capture such cities' welfare, given that it is noisy for nodes with low connectivity  and edge-weights~\cite{Ghoshal2011}. Furthermore, it does not capture an important feature, that is the maturity level of the cities from which travelers come to a given city. For instance, it stands to reason that for two developing cities with similar connectivity profiles, if one of them has visitors from mature cities, and the other only from developing cities, then the former may benefit more from the flow of intellectual and human capital from more developed regions.  

To account for this effect, we define the quantity 
\begin{equation}
    \langle S\rangle_i=\frac{\sum\limits_j W^I_{ji} S_j}{\sum\limits_j W^I_{ji}}\ ,
    \label{eq:theo_ind}
\end{equation}
where $S_j$ denotes the success/welfare of a given city $j$ (measured through the socioeconomic indicators) and ${\bf W^I}$ contains only international flows from $j$ to $i$.  We restrict ourselves to international flows, given that in our dataset, cities within a country have by and  large the same maturity levels. The quantity $\langle S \rangle_i$  can then be interpreted as the weighted average of the success of the international visitors to city $i$.  For instance if we take BHI as the measure of success, then this number will be high if most visitors are from Mature cities, and low if they are from Early Growth cities. Based on this we define an estimator for a city $i$'s success
\begin{equation}
    \hat S_i = PR_i \times \langle S\rangle_i^\gamma,
    \label{eq:est}
\end{equation}
where $PR_i$  is its PageRank and $\gamma$ is a tunable parameter that measures the extent to which the level of development of a city's neighbors plays any role in predicting its own development.  For $\gamma \simeq 0$, the node's network topological properties are the best indicator of its welfare, whereas for $\gamma \geq  0$ incorporating information about the development levels of its neighbors provides a better estimate.

\begin{table}
\caption{{\bf Role of mobility neighborhood for cities grouped by maturity} Values of $\gamma^{opt}$ for cities based on their maturity, optimizing the Spearman's correlation coefficient between the actual success $S_i$ (BHI) and the estimator $\hat S_i$ Eq.~\eqref{eq:est}. Size denotes the number of cities in each category. $\rho_s (\gamma)$ and $\rho_p (\gamma)$ are the Spearman's and Pearson's correlation coefficients for a given $\gamma$. Note that Pearson's correlation coefficients are computed by taking the logarithm of both $S$ and $\hat S$.}
\label{table1succ}
\begin{tabular}{ccccccc}
\hline
\bfseries Development-Stage &\bfseries Size &\bfseries $\gamma^{opt}$ & \bfseries $\rho_s(\gamma^{opt})$&\bfseries $\rho_p  (\gamma^{opt})$ &\bfseries $\rho_s(0)$ &\bfseries $\rho_p(0)$  \\
\hline
\begin{filecontents*}{mstable1.csv}
maturity,Size,gamma,Spearmanwithexponent,Spearmanwithoutexponent,pvalue_Spearman,Pearsonwithexponent,Pearsonwithoutexponent
Early Growth,64,1.302,0.572,0.378,0,0.586,0.414
Developing,44,0.108,0.885,0.877,0,0.875,0.873
Transitional,29,0.702,0.756,0.734,0,0.791,0.786
Mature,131,0.001,0.86,0.86,0,0.865,0.865
\end{filecontents*}
\csvreader[late after line=\\,]{mstable1.csv}{maturity=\Country,Size=\Size, gamma=\exponent, Spearmanwithexponent=\spwexp,
       Spearmanwithoutexponent=\spwoexp, Pearsonwithoutexponent=\pearsonwoexp, Pearsonwithexponent=\pearson}% 
       {\Country&\Size&\exponent&\spwexp&\pearson&\spwoexp&\pearsonwoexp}
 \\\hline \end{tabular}
\end{table}

Table~\ref{table1succ} contains the optimal value of the exponent $\gamma$ (see Figs. S10 and S11), maximizing the Spearman correlation $\rho_s(\gamma)$, between the estimator $\hat S_i$, and a city's actual level of success $S_i$ as quantified by the BHI which captures the city's  maturity level. A clear pattern emerges, whereby we find a decreasing $\gamma$ with increasing levels of development. The role of the average success of neighbors $\langle S \rangle$ is the most prominent for Early growth cities, and practically non-existent for Mature cities (see Fig. S12 for in-flow breakdown of Early Growth and Developing cities originating from national and international cities). Table~\ref{table2succ} contains the same information for cities categorized by Geographical Region. Mature cities in Western Europe and North America display the lowest values for $\gamma$, whereas regions such as Asia, Latin America and the Caribbean---that contain a mix of cities in terms of their development levels---show intermediate values for $\gamma$. The highest values are seen for the Middle East region and Central Europe that are home to fast rising cities. Interestingly Sub-Saharan Africa shows the same trend as Asia, though it is an outlier, as it seems neither the Pagerank nor the development levels of its neighbors are enough to account for explaining its own levels of development ($\rho_s(\gamma) =0.448)$ unlike for the other regions ($\rho_s(\gamma) \geq 0.763$).

 \begin{table}
\caption{{\bf Role of mobility neighborhood for cities grouped by geographic region}. All quantities same as in Tab.~\ref{table1succ}. }
 \label{table2succ}
\begin{tabular}{ccccccc}
\hline
\bfseries Region &\bfseries Size &\bfseries $\gamma^{opt}$ & \bfseries $\rho_s(\gamma^{opt})$&\bfseries $\rho_p  (\gamma^{opt})$ &\bfseries $\rho_s(0)$ &\bfseries $\rho_p(0)$  \\
\hline
\begin{filecontents*}{mstable2.csv}
Region,Size,gamma,Spearmanwithexponent,Spearmanwithoutexponent,pvalue_Spearman,Pearsonwithexponent,Pearsonwithoutexponent
CEE/CIS,27,2.973,0.781,0.651,0,0.664,0.751
MENA,18,2.532,0.835,0.733,2.00E-05,0.729,0.746
LATAM and Caribbean,30,0.672,0.822,0.716,0,0.886,0.883
Australasia,8,0.405,0.929,0.929,0.00086,0.949,0.942
Asia,36,0.377,0.885,0.856,0,0.873,0.866
Sub-Saharan Africa,16,0.331,0.448,0.387,0.12518,0.429,0.421
North America,64,0.053,0.862,0.858,0,0.873,0.774
Western Europe,69,-0.096,0.763,0.76,0,0.829,0.786
\end{filecontents*}
\csvreader[late after line=\\,]{mstable2.csv}{Region=\Country,Size=\Size, gamma=\exponent, Spearmanwithexponent=\spwexp,
       Spearmanwithoutexponent=\spwoexp, Pearsonwithoutexponent=\pearsonwoexp, Pearsonwithexponent=\pearson}% 
       {\Country&\Size&\exponent&\spwexp&\pearson&\spwoexp&\pearsonwoexp}
 \\\hline \end{tabular}
\end{table}

\subsection{The influence of success on mobility flows}

The enhanced ability to predict the true socioeconomic indicator of a city in the less developed categories by incorporating the maturity level of its neighborhood, suggests that perhaps success plays a role in how flows are allocated. To check for this, we construct a simple model for the allocation of the outgoing flows of a city among its destinations that accounts for both the success of the target destinations, as well as the distance between these targets and the source city. Suppose a city $i$ has a certain amount of total outflow $c_i$, which it can allocate towards different cities $j$ within some set of target cities $\partial_i$---its mobility network neighborhood. We assume that $i$ allocates its outgoing flow $c_i$ to neighbors $j\in \partial_i$ based on the success $S_j$ of city $j$, as well as the distance $d_{ij}$, so that $\tilde T_{ij}(\mu,\nu)\propto S_j^{\mu}d_{ij}^{\nu}$, where $\mu$ and $\nu$ are free parameters, and $\tilde T_{ij}(\mu,\nu)$ is the predicted flow from $i$ to $j$ from the model (whose dependence on $\mu$ and $\nu$ has been made explicit). More precisely, we have
\begin{align}
\label{eq:flow_model}
\tilde T_{ij}(\mu,\nu) = \frac{S_j^{\mu}d_{ij}^{\nu}}{\sum_{j\in \partial_i}S_j^{\mu}d_{ij}^{\nu}}\times c_i.    
\end{align}
If the success $S_j$ makes travel to city $j$ more attractive to city $i$, then one would expect $\mu\geq 0$. On the other hand, if the cities are far apart in distance $d_{ij}$, then it would make it more costly to travel from $i$ to $j$, and in that case $\nu \leq 0$. The ratio $|\frac{\mu}{\nu}|$ then determines the relative importance of these factors in determining $\tilde T_{ij}$. Given that $\tilde T_{ij}(\mu,\nu)$ is invariant to the scale of $S_j$ and $d_{ij}$, the relative importance of these measures can be compared directly using this ratio regardless of units. 

The outflow $c_i$ can be inferred from the raw (unweighted) origin-destination matrix $\bm{T}$ using the expression $c_i=\sum_{j\in \partial_i}T_{ij}$, where the neighborhood $\partial_i$ is restricted to the subset of cities $j$ that $i$ connects to that also have BHI metadata $S_j$. We then fit Eq.~\ref{eq:flow_model} to the true observed flows $T_{ij}$ to find the exponents $\hat\mu$ and $\hat\nu$ that optimize the Pearson correlation between the true and predicted flows $T_{ij}$ and $\tilde T_{ij}$ for connected city pairs $i,j$. This optimization is nontrivial but can be done approximately with a variety of methods, and here we choose a basin hopping algorithm \cite{wales2003energy}. Higher values of $|\hat\mu/\hat\nu|$ indicate that cities place high relative importance on success when allocating their outflows, and low values of $|\hat\mu/\hat\nu|$ indicate that city outflows are much more strongly influenced by distance than by success. To determine whether the city subgroup (development stage or geographical region) affects the influence of success on outgoing flows, we identify separate exponents $\{\hat\mu,\hat\nu\}$ for each subgroup by optimizing the Pearson correlation between $\tilde T_{ij}$ and $T_{ij}$ for all flows leaving cities $i$ in the subgroup.

We also determine whether or not the success covariate $S_j$ (BHI) significantly improves our predictive model of outflows over a baseline model, where only the distance between cities is considered. In this way, one can check if cities tend to place substantial importance on success when choosing how to allocate their flows. To do this, we identify the exponent $\nu_0$ that optimizes the correlation between the true and predicted flows while ignoring the success parameter ($\mu=0$). Since the model with $\mu$ has one more free parameter than the model with $\mu=0$, the fit between the true and predicted flows will always be better according to the Pearson correlation $\rho_p$. Thus for a fair comparison, the more complex model needs to be penalized using some model selection criteria. Here we opt for both the Akaike Information Criterion (AIC) and the Bayesian Information Criterion (BIC) \cite{hurvich1990impact}, which are derived from the log-likelihood of the model fit, as well as the number of data points in the sample. Table~S2 includes all model fit results for each category of cities. Applying both penalties indicate that the model performs significantly better with the inclusion of $\mu$ for all subgroups. We find that $\hat\mu>0$ and $\hat\nu<0$ for all city subgroups, indicating that cities allocate more flow to cities with higher BHI, and allocate less flow to cities that are farther away, though the extent to which  this happens depends on the level of development, with the effect most pronounced in Early Growth and Developing cities, and least so in Mature and Transitional cities.  

\begin{figure}[t!]
	\includegraphics[width=\textwidth]{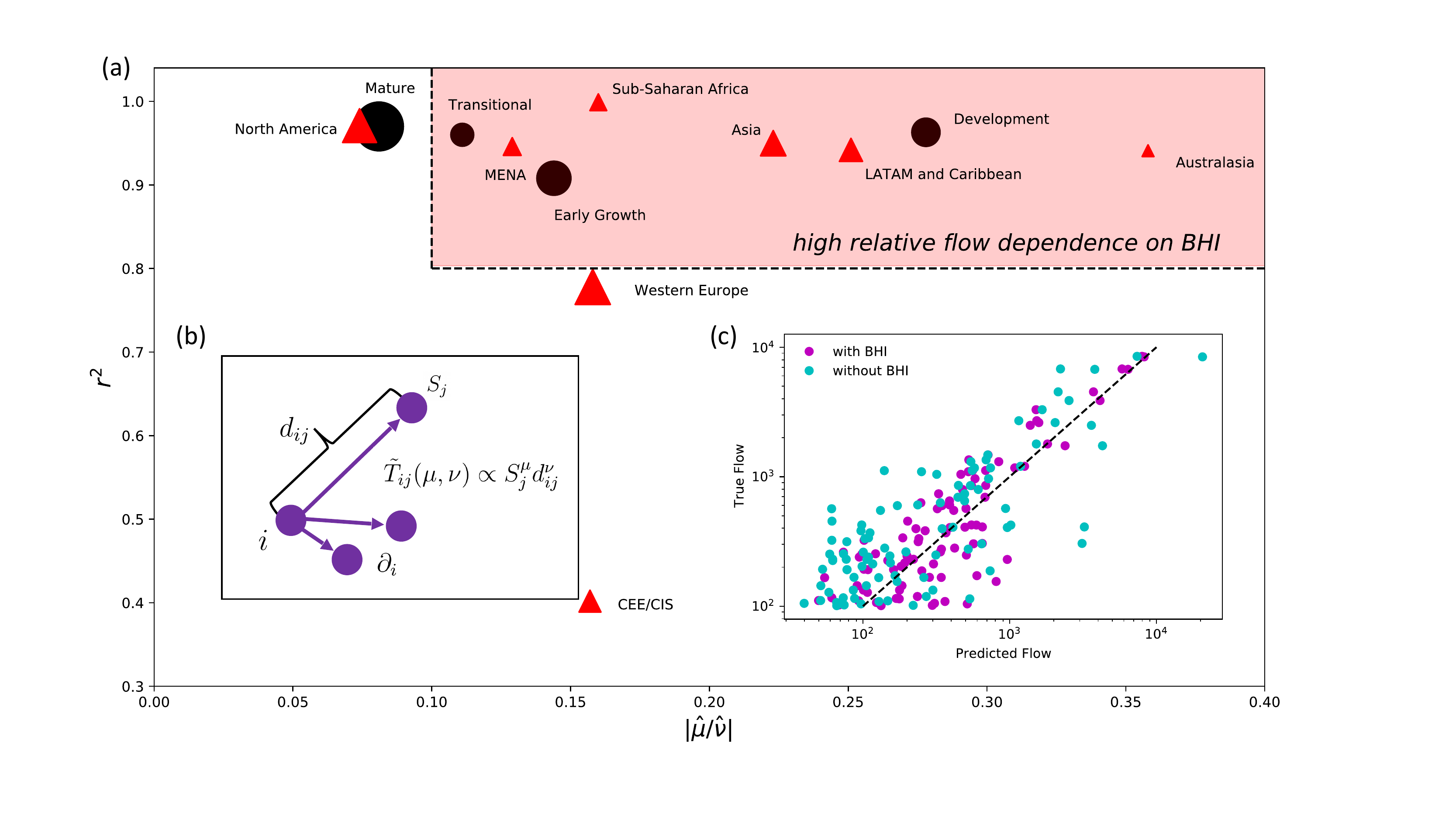}\\
	\caption{{\bf Interplay between benefit (success) and cost (distance)}\textbf{a} $|\hat\mu/\hat\nu|$ and coefficient of determination $r^2$ for all city subgroups considered, using the model in Eq.~\ref{eq:flow_model} to fit the outflows of all cities in the subgroup. Black circles indicate development subgroups, and red triangles indicate geographical subgroups, while the size of the marker is proportional to the number of cities in the subgroup. Cities with $r^2>0.8$ and $|\hat\mu/\hat\nu|>0.1$ are highlighted as having high relative flow dependence on BHI due to high model correlations and contributions from $S_j^\mu$. All fits demonstrated significant improvement through the inclusion of the $S_j^\mu$ term, as evidenced by the corresponding AIC and BIC values (Table.~\ref{sitab:modelsummary} in SI). \textbf{b} Model schematic, showing the variables involved in Eq.~\ref{eq:flow_model} around a central node $i$ for which outflows $\tilde T_{ij}$ are being predicted. \textbf{c} Example predicted and true outflows from cities in the Australasia subgroup, showing significant improvement from including the BHI in the model fit. Only flows with $T_{ij}>100$ are displayed, for clearer visualization, and the line of equality is shown for reference.}
	\label{fig:flow_model}	
\end{figure}
\par

In Fig.~\ref{fig:flow_model}(a), the relative magnitude $|\hat\mu/\hat\nu|$ of the coefficients $\hat\mu$ and $\hat\nu$ is plotted along with the coefficient of determination $r^2$ for each city subcategory. As discussed, the ratio $|\hat\mu/\hat\nu|$ quantifies the relative importance of success $S_j$ versus distance $d_{ij}$ in determining how a city $i$ allocates its flow $c_i$.  According to the model fits, in general developing cities place more importance on success than more developed cities, as indicated by the shaded red box. More specifically, cities in the Developing, Early Growth, and Transitional subgroup place higher relative priority on the success of the cities they allocate flows to as compared to cities in the Mature subgroup. We also see a geographic dependence with North America and (Western and Eastern) Europe placing relatively lower importance on BHI and having weaker model fits in general. This suggests that other factors are more important in determining the corresponding flows (perhaps consistent with these regions having many developed cities). As insets, we illustrate the variables used in the flow model (Fig.~\ref{fig:flow_model}(b)), as well as show an example of the improved weight prediction after the inclusion of BHI for the Australasia subgroup (Fig.~\ref{fig:flow_model}(c)). In this third panel, we plot the flow as predicted by Eq.~\ref{eq:flow_model}---with $\hat\mu$ and $\hat\nu$ as the model parameters---along the x-axis, and the true observed flow along the y-axis. The weights condense significantly around the line of equality when including $S_j$ as a co-variate, as compared to using only $d_{ij}$. 

These results indicate that cities in earlier stages of development or in more developing regions place greater importance on their neighbors' success when considering their mobility outflows. These results compliment our previous analysis by suggesting that perhaps such cities strategically allocate their outflows in order to maximize the benefits they receive by connecting with more successful cities. On the other hand, for developed cities outflows are more strongly influenced by travel distance.          

\section{Discussion}   

We have presented here, a comprehensive analysis on the role of inter-city global mobility flows apparent in driving the welfare and success of urban areas. At the macroscopic level, we report the existence of a tightly connected core of cities that form a rich-club dominating mobility flow in the network. Cities in this core tend to have higher levels of development, and are primarily located in Western Europe and North America. For these cities, their development-level is well predicted by their network topological properties, in particular centrality measures such as PagerRank. On the other hand developing and early cities are located in the peripheral layers and are scattered across multiple regions (Asia, Eastern Europe, Latin America, Middle East  and Sub-Saharan Africa).  Their level of development is partially connected to their network centrality in some cases, and very weakly so in others, in particular in Africa.

\par
For those regions, whose success is weakly connected to their centrality, we find that adding information on their mobility neighborhood (that is the average development level of the cities they are connected to) leads to a marked increase in correlations with their socioeconomic indicators. This effect is less pronounced for more developed cities, whose socioeconomic indicators are connected primarily with their network centrality. Once again Sub-Saharan Africa is an outlier, its development being weakly connected to network properties and the socioeconomics of their mobility neighborhood.  To check whether the connection of cities to other cities is indeed influenced by their  development-level we propose a simple model that disentangles the relevant aspects influencing the success/welfare of urban areas. The model assumes that the outflow of cities is distributed among destinations in proportion to their relative benefit (maturity) and cost (distance). The predicted outflows are compared to the empirical outflows to determine the contribution of each individual component of the model. We find that all cities incorporate the development level of the cities they connect to, and are less inclined to connect to cities at great distance. However, the former is much more pronounced in developing regions as compared to developed regions. 

\par
Taken together, our analysis indicates that the welfare of cities and the inter-urban mobility network are strongly correlated, but that this correlation depends on their level of maturity. The global urban ecosystem appears to be a combination of a well-established set of core urban areas whose interconnections describe their performance according to network science principles, and a subset of developing cities whose welfare is influenced by the extent of connections to the core. While we do not have data on individual mobility-flows the results are akin to that seen in~\cite{Fraiberger2018}, where artists climb the ladder, as it were, by strategically positioning themselves in the network of galleries exhibiting their work. 

\section{Limitations}
These results should be interpreted in light of several important limitations. First, the Google mobility data is limited to smartphone users who have opted in to Google Location History feature, which is off by default. These data may not be representative of the population as whole, and may vary by location. Importantly, these limited data are only viewed through the lens of differential privacy algorithms, specifically designed to protect user anonymity and obscure fine detail. Moreover, comparisons across rather than within locations are only descriptive since these regions can differ in substantial ways.
\section{Data Availability}
The Google Aggregated Mobility Research Data-set used for this study is available with permission from Google LLC. Links to all other data sources have been provided in the manuscript.
\par
\par
\noindent {\bf Acknowledgments}

We thank A. Bassolas for useful discussions on the manuscript. G.G. and  S.M. acknowledge partial support from the U. S. Army Research Office under grant number W911NF-18-1-0421 and National Science Foundation under Grant No 2029095. A.A. acknowledges financial support from Spanish Ministerio de Ciencia e Innovaci\'on (grant PGC2018-094754-BC21), Generalitat de Catalunya (grant No. 2017SGR-896 and 2020PANDE00098), and Universitat Rovira i Virgili (grant No. 2019PFR-URVB2-41). A.A. also acknowledges support from Generalitat de Catalunya ICREA Academia, and the James S. McDonnell Foundation (grant 220020325). D.S.P. and J.G.G. acknowledge financial support from Spanish Ministerio de Ciencia e Innovaci\'on (projects FIS2017-87519-P and PID2020-113582GB-I00), from the Departamento de Industria e Innovaci\'on del Gobierno de Arag\'on y Fondo Social Europeo (FENOL group E-19), and from Fundaci\'on Ibercaja and Universidad de Zaragoza (grant 224220). A.K. is supported by the National Defense Science and Engineering Graduate Fellowship
through the Department of Defense.

\bibliographystyle{naturemag}
\bibliography{references}
\end{document}

\makeatletter
\renewcommand{\maketitle}{\bgroup\setlength{\parindent}{0pt}
	\begin{flushleft}
{\huge Supporting Information}\\[0.3cm]
		{\large\textbf \@title}
		
		\@author
		
	\end{flushleft}\egroup
}
\makeatother
\title{The impact of inter-city mobility on urban welfare}

\author{S. Mimar et al.}                         
%\affil[1]{\RochesterP}
% \affil[2]{\France}
%\affil[3]{\france}
%\affil[4]{\RochesterC}

\date{}

\maketitle

\renewcommand{\contentsname}{Table of Contents}
\tableofcontents
\makeatletter
\let\toc@pre\relax
\makeatother 

\listoffigures
\listoftables

\clearpage
\doublespacing
\section{Global cities with socioeconomic indicators}

\begin{figure}[ht!]	
		\centering
		\includegraphics[width=0.90\linewidth]{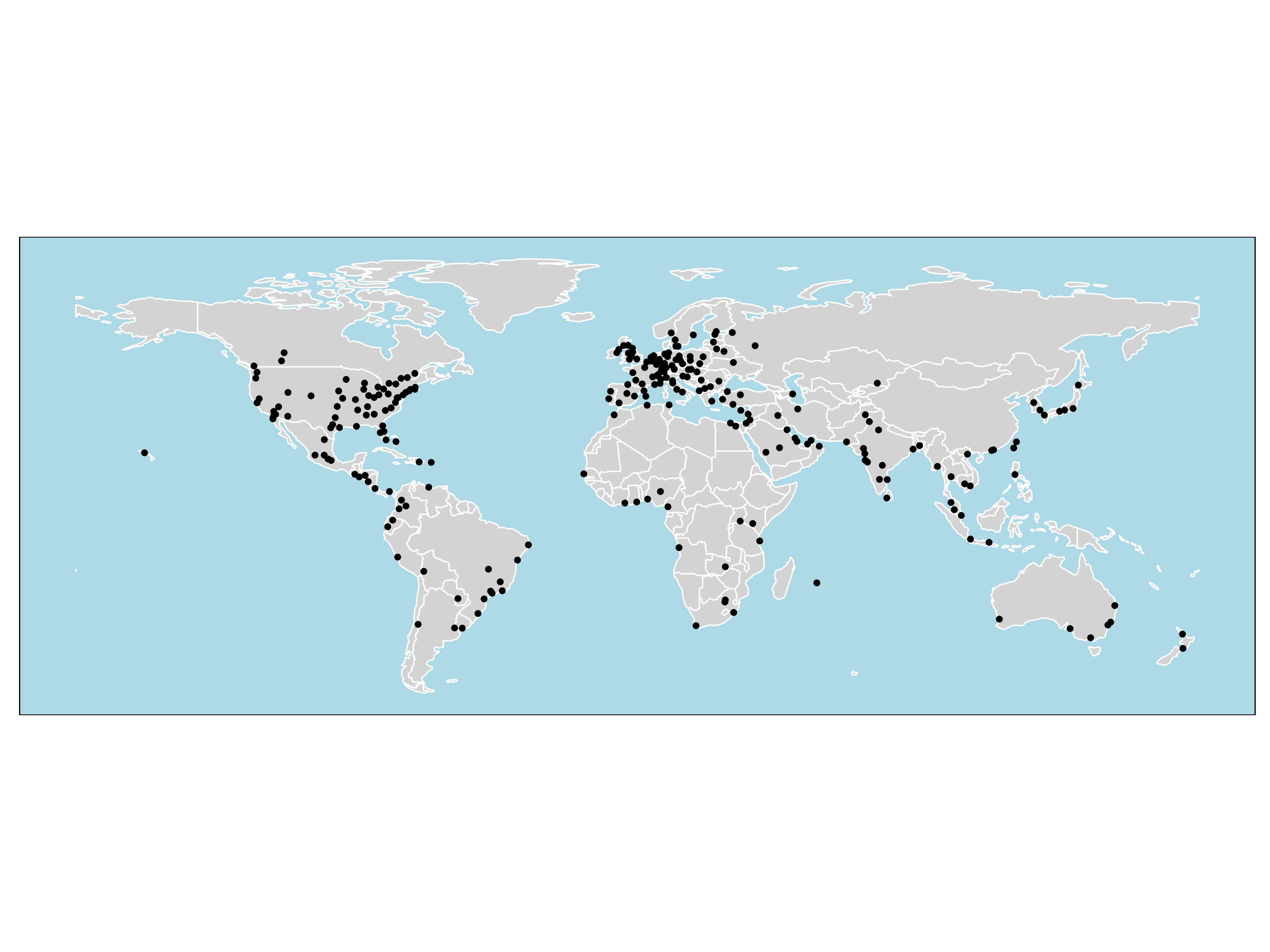}
		\vspace{-1in}
	\caption[Global cities selected for the study]{Global cities selected for the study}
		\label{sifig:all_cities}
\end{figure}

\begin{figure}[ht!]	
		\centering
		\includegraphics[width=0.90\linewidth]{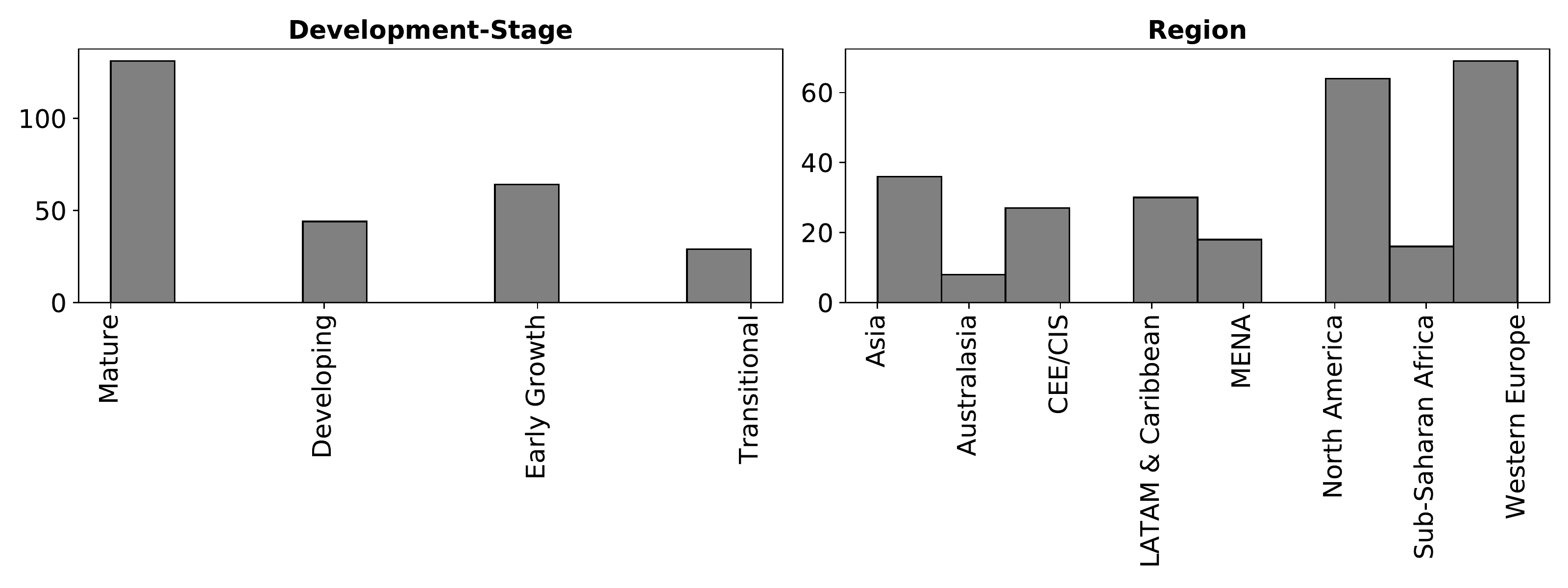}
	\caption[Histograms of cities in subcategories]{Histograms of cities in subcategories}
		\label{sifig:hist_all_cities}
\end{figure}

In this study, we use 268 urban areas that have socioeconomic indicators (BHI, GDP, Total Real Estate Investment,Total Cross boundary investment), published by JLL and illustrated in the map below \ref{sifig:all_cities} and listed in Table ~\ref{sitab:list_cities}. In addition to the socioeconomic indicators, cities are categorized with respect to their development level and geographic location. In Fig. \ref{sifig:hist_all_cities}, we show number of urban areas in each subcategory.
\vspace{-0.1in}
\newpage
\clearpage

\footnotesize{
\begin{center}
\begin{filecontents*}{smtable1.csv}
a,b,c,d
Surat,India,Asia,Early Growth
Tokyo,Japan,Asia,Mature
Lahore,Pakistan,Asia,Early Growth
Nagoya,Japan,Asia,Transitional
Hong Kong,Hong Kong,Asia,Mature
Pune,India,Asia,Developing
Singapore,Singapore,Asia,Mature
Osaka,Japan,Asia,Mature
Jakarta,Indonesia,Asia,Developing
Seoul,South Korea,Asia,Transitional
Ahmedabad,India,Asia,Early Growth
Dhaka,Bangladesh,Asia,Early Growth
Penang-Georgetown,Malaysia,Asia,Developing
Colombo,Sri Lanka,Asia,Early Growth
Yangon,Myanmar,Asia,Early Growth
Hanoi,Vietnam,Asia,Early Growth
Bangalore,India,Asia,Developing
Sapporo,Japan,Asia,Transitional
Chennai,India,Asia,Developing
Kaohsiung,Taiwan,Asia,Developing
Karachi,Pakistan,Asia,Early Growth
Ho Chi Minh City,Vietnam,Asia,Early Growth
Kolkata,India,Asia,Developing
Taipei,Taiwan,Asia,Transitional
Busan,South Korea,Asia,Developing
Surabaya,Indonesia,Asia,Early Growth
Fukuoka,Japan,Asia,Transitional
Manila,Philippines,Asia,Developing
Kuala Lumpur,Malaysia,Asia,Transitional
Delhi,India,Asia,Developing
Hyderabad,India,Asia,Developing
Bangkok,Thailand,Asia,Developing
Islamabad,Pakistan,Asia,Early Growth
Mumbai,India,Asia,Developing
Macau,Macau,Asia,Developing
Phnom Penh,Cambodia,Asia,Early Growth
Brisbane,Australia,Australasia,Mature
Wellington,New Zealand,Australasia,Mature
Perth,Australia,Australasia,Mature
Adelaide,Australia,Australasia,Mature
Newcastle,Australia,Australasia,Mature
Sydney,Australia,Australasia,Mature
Auckland,New Zealand,Australasia,Mature
Melbourne,Australia,Australasia,Mature
Tirana,Albania,CEE/CIS,Early Growth
Istanbul,Turkey,CEE/CIS,Developing
Skopje,Macedonia,CEE/CIS,Early Growth
Minsk,Belarus,CEE/CIS,Early Growth
Poznan,Poland,CEE/CIS,Developing
Antalya,Turkey,CEE/CIS,Early Growth
Moscow,Russia,CEE/CIS,Transitional
Almaty,Kazakhstan,CEE/CIS,Early Growth
Izmir,Turkey,CEE/CIS,Early Growth
Ankara,Turkey,CEE/CIS,Developing
Baku,Azerbaijan,CEE/CIS,Early Growth
Prague,Czech Republic,CEE/CIS,Transitional
Kiev,Ukraine,CEE/CIS,Developing
Vilnius,Lithuania,CEE/CIS,Transitional
Sofia,Bulgaria,CEE/CIS,Developing
Riga,Latvia,CEE/CIS,Transitional
Krakow,Poland,CEE/CIS,Developing
Zagreb,Croatia,CEE/CIS,Early Growth
Tallinn,Estonia,CEE/CIS,Transitional
Bucharest,Romania,CEE/CIS,Developing
St Petersburg,Russia,CEE/CIS,Developing
Wroclaw,Poland,CEE/CIS,Developing
Warsaw,Poland,CEE/CIS,Transitional
Budapest,Hungary,CEE/CIS,Transitional
Bratislava,Slovakia,CEE/CIS,Transitional
Ljubljana,Slovenia,CEE/CIS,Developing
Belgrade,Serbia,CEE/CIS,Early Growth
Porto Alegre,Brazil,LATAM/Caribbean,Early Growth
La Paz,Bolivia,LATAM/Caribbean,Early Growth
Asuncion,Paraguay,LATAM/Caribbean,Early Growth
Campinas,Brazil,LATAM/Caribbean,Early Growth
Rio de Janeiro,Brazil,LATAM/Caribbean,Developing
Bogota,Colombia,LATAM/Caribbean,Developing
Santiago,Chile,LATAM/Caribbean,Transitional
San Jose,Costa Rica,LATAM/Caribbean,Developing
Medellin,Colombia,LATAM/Caribbean,Early Growth
Curitiba,Brazil,LATAM/Caribbean,Early Growth
Belo Horizonte,Brazil,LATAM/Caribbean,Developing
Sao Paulo,Brazil,LATAM/Caribbean,Developing
Cali,Colombia,LATAM/Caribbean,Early Growth
Panama City,Panama,LATAM/Caribbean,Early Growth
Quito,Ecuador,LATAM/Caribbean,Early Growth
Nassau,Bahamas,LATAM/Caribbean,Transitional
Guatemala City,Guatemala,LATAM/Caribbean,Early Growth
Montevideo,Uruguay,LATAM/Caribbean,Developing
Tegucigalpa,Honduras,LATAM/Caribbean,Early Growth
Caracas,Venezuela,LATAM/Caribbean,Early Growth
Buenos Aires,Argentina,LATAM/Caribbean,Developing
Santo Domingo,Dominican Republic,LATAM/Caribbean,Early Growth
Guayaquil,Ecuador,LATAM/Caribbean,Early Growth
Recife,Brazil,LATAM/Caribbean,Early Growth
Managua,Nicaragua,LATAM/Caribbean,Early Growth
San Juan,Puerto Rico,LATAM/Caribbean,Developing
Salvador,Brazil,LATAM/Caribbean,Early Growth
San Salvador,El Salvador,LATAM/Caribbean,Early Growth
Brasilia,Brazil,LATAM/Caribbean,Developing
Lima,Peru,LATAM/Caribbean,Early Growth
Dubai,United Arab Emirates,MENA,Developing
Baghdad,Iraq,MENA,Early Growth
Jeddah,Saudi Arabia,MENA,Early Growth
Doha,Qatar,MENA,Developing
Amman,Jordan,MENA,Early Growth
Abu Dhabi,United Arab Emirates,MENA,Developing
Algiers,Algeria,MENA,Early Growth
Cairo,Egypt,MENA,Early Growth
Alexandria,Egypt,MENA,Early Growth
Kuwait City,Kuwait,MENA,Early Growth
Tunis,Tunisia,MENA,Early Growth
Tehran,Iran,MENA,Early Growth
Casablanca,Morocco,MENA,Developing
Manama,Bahrain,MENA,Developing
Riyadh,Saudi Arabia,MENA,Early Growth
Tel Aviv,Israel,MENA,Transitional
Beirut,Lebanon,MENA,Early Growth
Muscat,Oman,MENA,Early Growth
Columbus,USA,North America,Mature
Las Vegas,USA,North America,Mature
Austin,USA,North America,Mature
Raleigh-Durham,USA,North America,Mature
Kansas City,USA,North America,Mature
Milwaukee,USA,North America,Mature
Salt Lake City,USA,North America,Mature
New York,USA,North America,Mature
Pittsburgh,USA,North America,Mature
Memphis,USA,North America,Mature
Quebec City,Canada,North America,Mature
Puebla,Mexico,North America,Early Growth
Stamford- Bridgeport-Norwalk,USA,North America,Mature
Rochester,USA,North America,Mature
Birmingham,USA,North America,Mature
Tulsa,USA,North America,Mature
Queretaro,Mexico,North America,Developing
Omaha,USA,North America,Mature
Providence,USA,North America,Mature
Jacksonville,USA,North America,Mature
Tijuana,Mexico,North America,Developing
Indianapolis,USA,North America,Mature
Riverside-San Bernardino,USA,North America,Mature
New Orleans,USA,North America,Mature
Richmond,USA,North America,Mature
Edmonton,Canada,North America,Mature
Ottawa,Canada,North America,Mature
Sacramento,USA,North America,Mature
San Antonio,USA,North America,Mature
Nashville,USA,North America,Mature
Honolulu,USA,North America,Mature
Hartford,USA,North America,Mature
Cincinnati,USA,North America,Mature
Guadalajara,Mexico,North America,Developing
Baltimore,USA,North America,Mature
Boston,USA,North America,Mature
Cleveland,USA,North America,Mature
Tampa,USA,North America,Mature
St Louis,USA,North America,Mature
Dallas,USA,North America,Mature
Monterrey,Mexico,North America,Developing
San Francisco,USA,North America,Mature
Atlanta,USA,North America,Mature
Vancouver,Canada,North America,Mature
Washington,USA,North America,Mature
Calgary,Canada,North America,Mature
Mexico City,Mexico,North America,Developing
Seattle,USA,North America,Mature
Charlotte,USA,North America,Mature
Toronto,Canada,North America,Mature
Miami,USA,North America,Mature
Houston,USA,North America,Mature
San Jose,USA,North America,Mature
Philadelphia,USA,North America,Mature
Detroit,USA,North America,Mature
Orlando,USA,North America,Mature
San Diego,USA,North America,Mature
Phoenix,USA,North America,Mature
Portland,USA,North America,Mature
Los Angeles,USA,North America,Mature
Denver,USA,North America,Mature
Montreal,Canada,North America,Mature
Chicago,USA,North America,Mature
Minneapolis,USA,North America,Mature
Cape Town,South Africa,Sub-Saharan Africa,Transitional
Pretoria,South Africa,Sub-Saharan Africa,Transitional
Durban,South Africa,Sub-Saharan Africa,Transitional
Lagos,Nigeria,Sub-Saharan Africa,Early Growth
Dar es Salaam,Tanzania,Sub-Saharan Africa,Early Growth
Dakar,Senegal,Sub-Saharan Africa,Early Growth
Accra,Ghana,Sub-Saharan Africa,Early Growth
Port Louis,Mauritius,Sub-Saharan Africa,Developing
Abidjan,Cote D'Ivoire,Sub-Saharan Africa,Early Growth
Kampala,Uganda,Sub-Saharan Africa,Early Growth
Lusaka,Zambia,Sub-Saharan Africa,Early Growth
Nairobi,Kenya,Sub-Saharan Africa,Early Growth
Luanda,Angola,Sub-Saharan Africa,Early Growth
Douala,Cameroon,Sub-Saharan Africa,Early Growth
Johannesburg,South Africa,Sub-Saharan Africa,Transitional
Abuja,Nigeria,Sub-Saharan Africa,Early Growth
Nuremburg,Germany,Western Europe,Mature
Malmo,Sweden,Western Europe,Mature
Florence,Italy,Western Europe,Mature
Lille,France,Western Europe,Mature
Dresden,Germany,Western Europe,Mature
Copenhagen,Denmark,Western Europe,Mature
Edinburgh,United Kingdom,Western Europe,Mature
Utrecht,Netherlands,Western Europe,Mature
Zurich,Switzerland,Western Europe,Mature
Lyon,France,Western Europe,Mature
Hamburg,Germany,Western Europe,Mature
Lausanne,Switzerland,Western Europe,Mature
Brussels,Belgium,Western Europe,Mature
Rome,Italy,Western Europe,Mature
Hannover,Germany,Western Europe,Mature
Turin,Italy,Western Europe,Mature
Vienna,Austria,Western Europe,Mature
Dublin,Ireland,Western Europe,Mature
Naples,Italy,Western Europe,Transitional
Lisbon,Portugal,Western Europe,Mature
Dusseldorf,Germany,Western Europe,Mature
Newcastle-upon-Tyne,United Kingdom,Western Europe,Mature
Oslo,Norway,Western Europe,Mature
Geneva,Switzerland,Western Europe,Mature
Rotterdam,Netherlands,Western Europe,Mature
Belfast,United Kingdom,Western Europe,Mature
Marseilles,France,Western Europe,Mature
Luxembourg,Luxembourg,Western Europe,Mature
Manchester,United Kingdom,Western Europe,Mature
Seville,Spain,Western Europe,Transitional
Stuttgart,Germany,Western Europe,Mature
Leipzig,Germany,Western Europe,Mature
Porto,Portugal,Western Europe,Transitional
Gothenburg,Sweden,Western Europe,Mature
Leeds,United Kingdom,Western Europe,Mature
Nantes,France,Western Europe,Mature
Birmingham,United Kingdom,Western Europe,Mature
Valencia,Spain,Western Europe,Transitional
Athens,Greece,Western Europe,Transitional
Helsinki,Finland,Western Europe,Mature
Bilbao,Spain,Western Europe,Transitional
Toulouse,France,Western Europe,Mature
Essen,Germany,Western Europe,Mature
Basel,Switzerland,Western Europe,Mature
Liverpool,United Kingdom,Western Europe,Mature
Bonn,Germany,Western Europe,Mature
Bremen,Germany,Western Europe,Mature
Madrid,Spain,Western Europe,Mature
Glasgow,United Kingdom,Western Europe,Mature
Milan,Italy,Western Europe,Mature
Nicosia,Cyprus,Western Europe,Transitional
Mannheim,Germany,Western Europe,Mature
Paris,France,Western Europe,Mature
Antwerp,Belgium,Western Europe,Mature
Bordeaux,France,Western Europe,Mature
Cologne,Germany,Western Europe,Mature
Bologna,Italy,Western Europe,Mature
Palma de Mallorca,Spain,Western Europe,Transitional
The Hague,Netherlands,Western Europe,Mature
Stockholm,Sweden,Western Europe,Mature
Bristol,United Kingdom,Western Europe,Mature
Berlin,Germany,Western Europe,Mature
London,United Kingdom,Western Europe,Mature
Amsterdam,Netherlands,Western Europe,Mature
Munich,Germany,Western Europe,Mature
Strasbourg,France,Western Europe,Mature
Nice,France,Western Europe,Mature
Frankfurt,Germany,Western Europe,Mature
Barcelona,Spain,Western Europe,Mature
\end{filecontents*}
\csvreader[
longtable=|c|c|c|c|,
table head =\caption{List of global cities with their Geographic Region and Development-Stage.}\\ \hline\textbf{City}&\textbf{Country}&\textbf{Region}&\textbf{Development-Stage}\\\hline,
before table=\rowcolors{2}{white}{lightgray},
late after last line=\\\hline]{smtable1.csv}{a=\a,b=\b,c=\c,d=\d}{\a&\b&\c&\d}
\label{sitab:list_cities}
\end{center}

}
\clearpage
\section{Comparison of PageRank and strength correlations with BHI}
\begin{figure}[h]
\centering
		\includegraphics[width=0.9\linewidth]{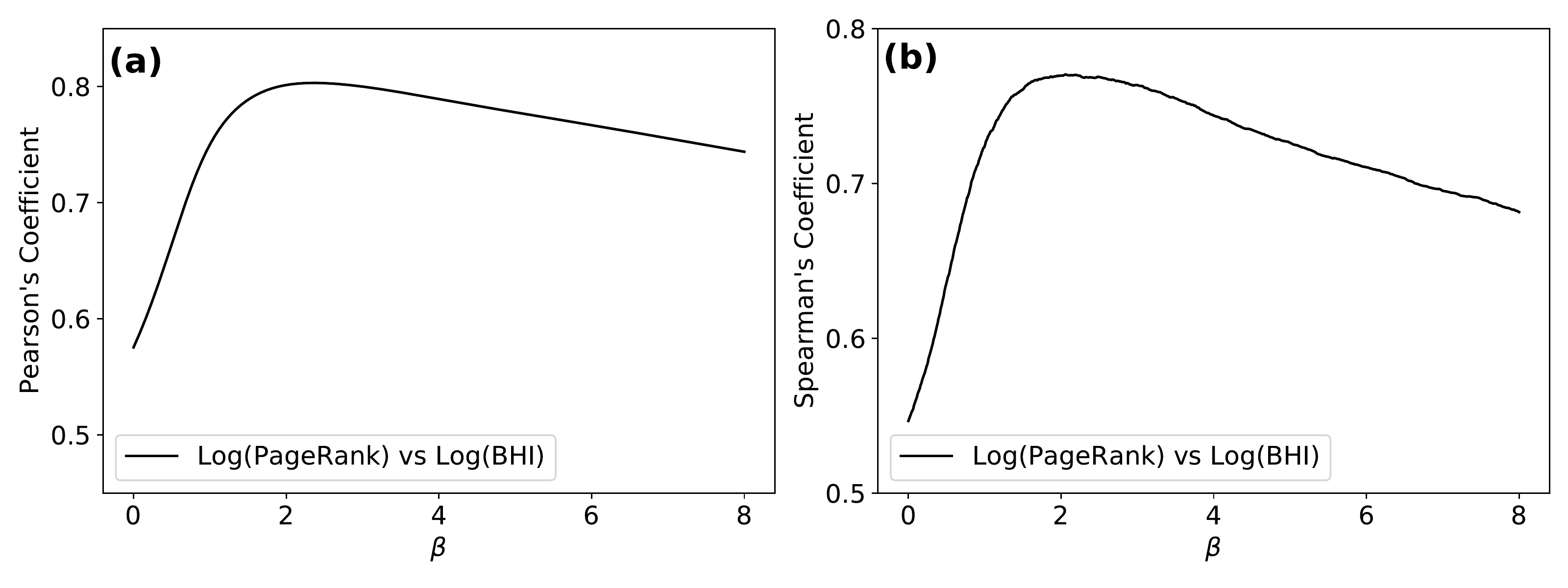}
	\caption[Dependence of correlation coefficients on the weight exponent $\beta$]{Pearson's  in \textbf{a} and  Spearman's in \textbf{b} correlation coefficient for Pagerank vs BHI as the distance exponent $\beta$ is varied when computing the weights $W_{ij}=T_{ij}d_{ij}^\beta$. The curves yield a peak at $\beta = 2.5$ where the correlation coefficients are maximized.}
	
	\label{sifig:p_s_v_exp}
\end{figure}

\begin{figure}[ht!]	
	\centering
	\includegraphics[width=0.5\linewidth]{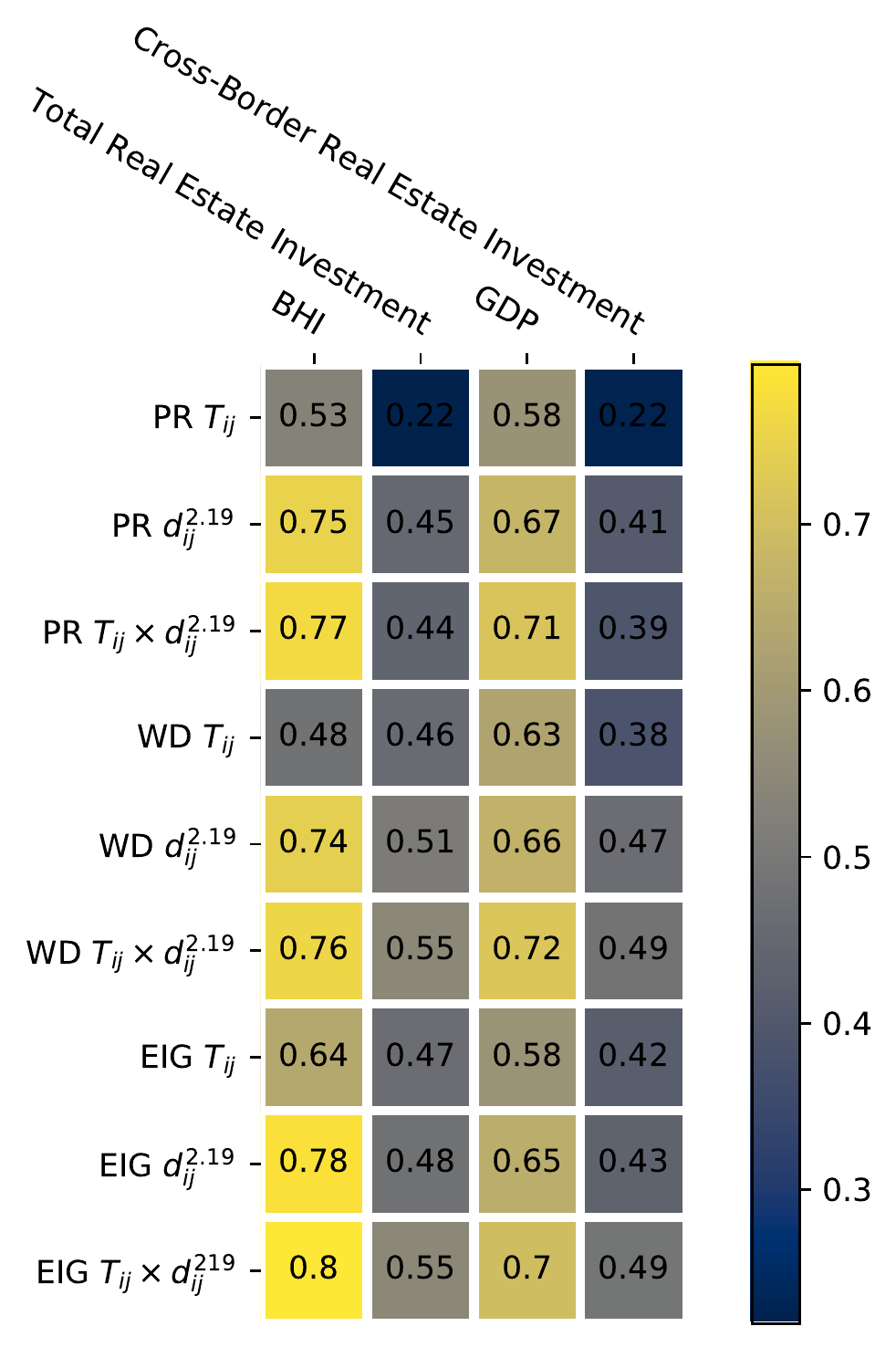}
	\caption[Global Correlation Summary ]{Correlation summary of mobility network centrality measures for clobal cities in the dataset, namely Pagerank (PR), Weighted-Degree (WD) and eigenvector centrality (EIG) versus socioeconomic indicators: BHI, Total real Estate investment, GDP and Corss-Border Real Estate Investment. Metrics are computed with various edge-weights to show the performance of flow ($T_{ij}$), distance $d_{ij}$ and flow $\times$ distance ($T_{ij}d_{ij}$). The correlation metric used here is the Spearman's coefficient $\rho_s$.}
\label{sifig:summary_global}
\end{figure}
\clearpage
\newpage
\section{Categorical comparison}
\label{section:catcom}
\begin{figure}[ht!]	
		\centering
		\includegraphics[width=0.83\linewidth]{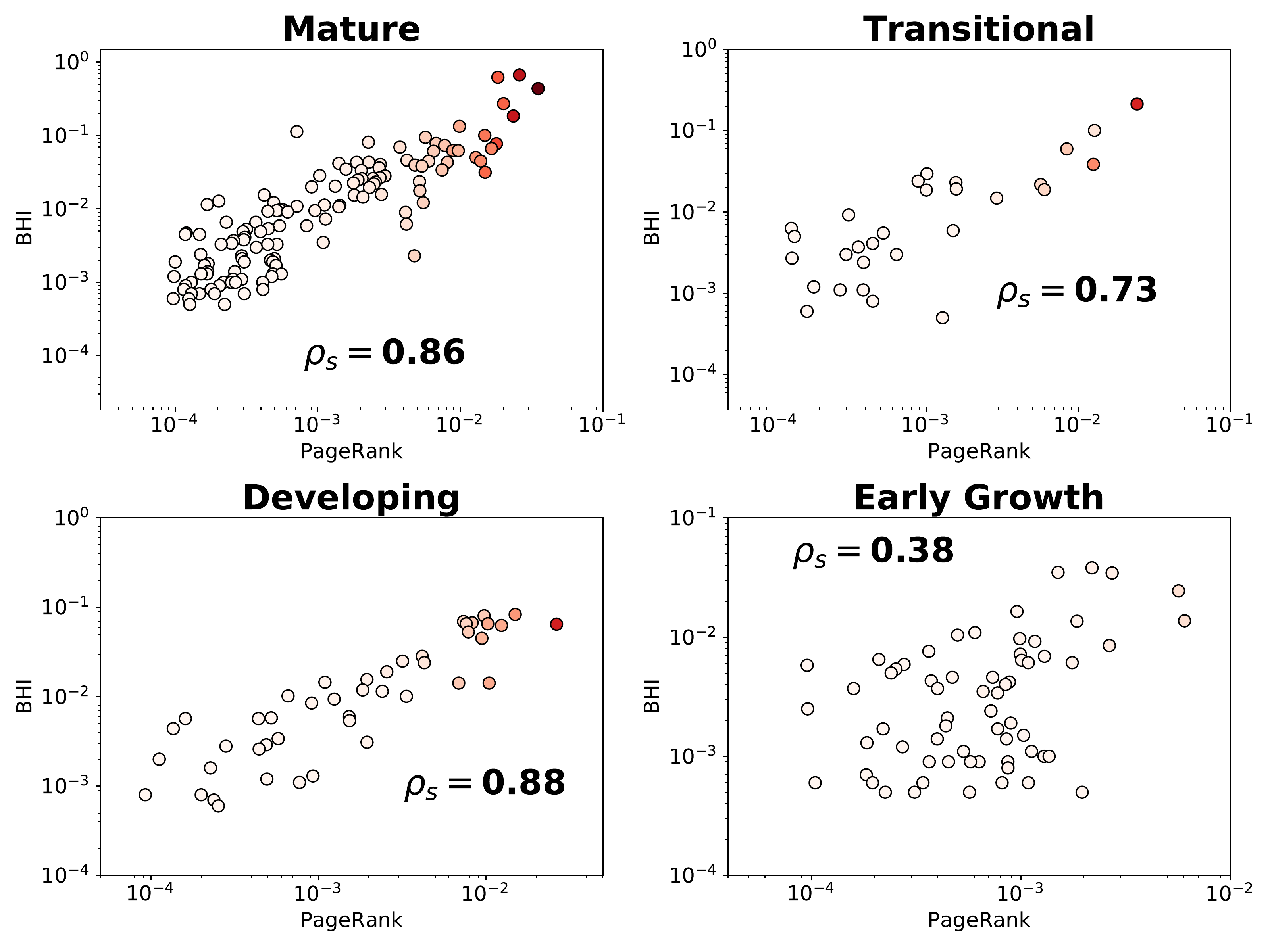}
	\caption[Pagerank vs BHI at development level]{Pagerank vs BHI when cities are grouped with respect to their development level. Spearman's correlation in each panel shows association level of the socioeconomic indicator and the network centrality measure. Colors represent the strength of each city with dark red indicating high Weighted-Degree( WD).}
		\label{sifig:devlev}
\end{figure}

\begin{figure}[ht!]	
		\centering
		\includegraphics[width=0.83\linewidth]{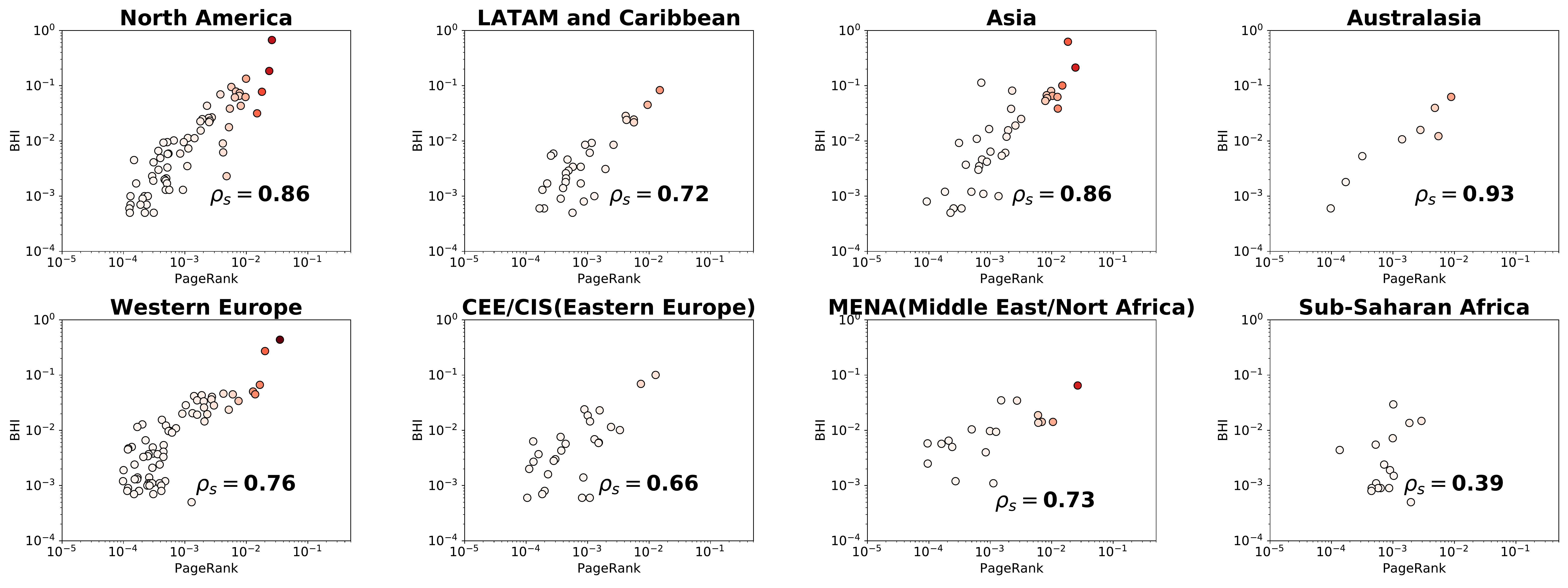}
	\caption[Pagerank vs BHI at geographic region level]{Pagerank vs BHI when cities are grouped according to geographic regions. Spearman's correlation in each panel shows association level of the socioeconomic indicator and the network centrality measure.Colors represent the strength of each city with dark red indicating high Weighted-Degree(WD).}
		\label{sifig:geo}
\end{figure}

\newpage

\begin{figure}[ht!]	
	\centering
	\includegraphics[width=1.0\linewidth]{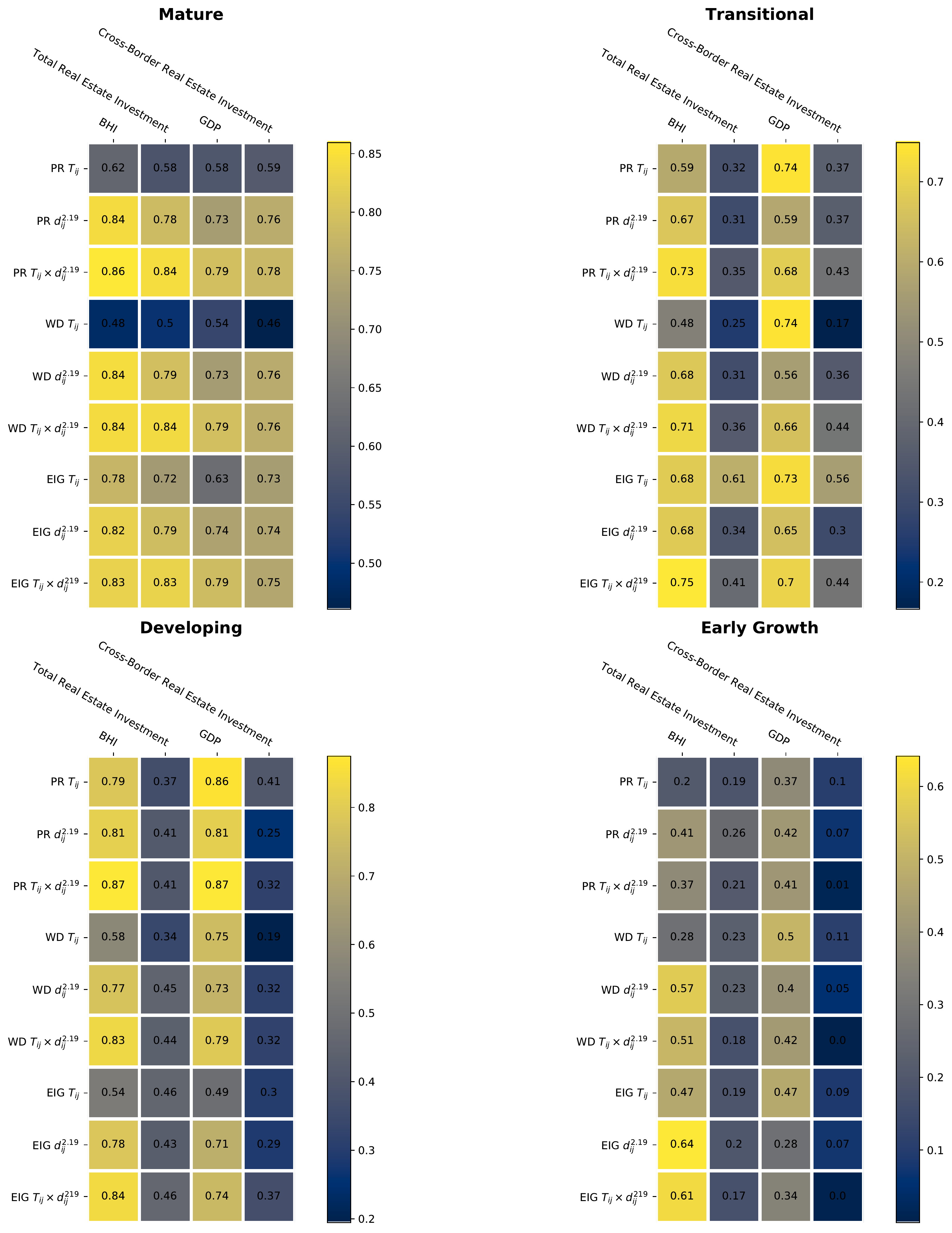}
	\caption[Correlation Summary at Development Level]{Correlation summary of mobility network centrality measures, namely Pagerank (PR), Weighted-Degree (WD) and eigenvector centrality (EIG) versus socioeconomic indicators: BHI, Total real Estate investment, GDP and Corss-Border Real Estate Investment. Metrics are computed with various edge-weights to show the performance of flow ($T_{ij}$), distance $d_{ij}$ and flow $\times$ distance ($T_{ij}d_{ij}$). The correlation metric used here is the Spearman's coefficient $\rho_s$.}
\label{sifig:summary}
\end{figure}

\newpage
\section{Core-Periphery Structure}
\label{section:core}

\begin{figure}[ht!]	
		\centering
		\includegraphics[width=1.0\linewidth]{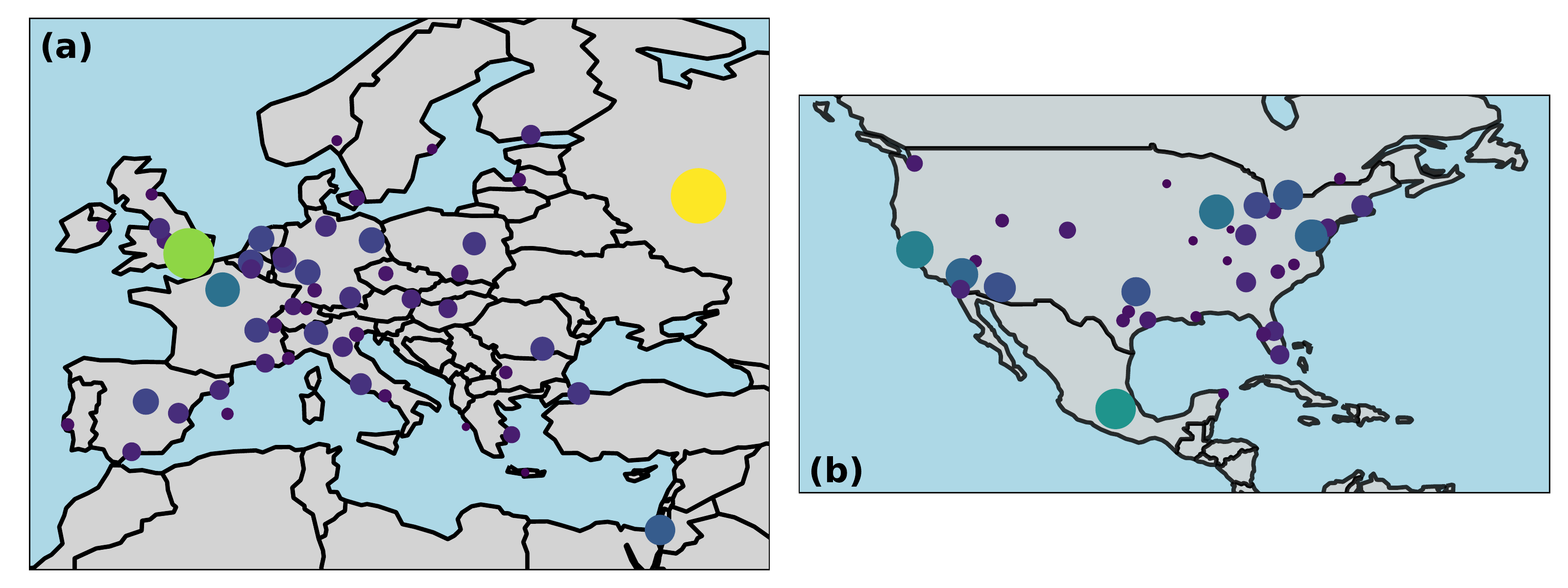}
	\caption[Core cities]{Two cores of the mobility network \textbf{a} coreness = $69$ that consists mostly Western European cities, \textbf{b} coreness = $74$ where the clusters are formed by North American cities. Size and colors of the nodes are proportional to their PageRank (from light to dark, large to small for high to low PageRank).}
		\label{sifig:core_clusters_map}
\end{figure}

\begin{figure}[ht!]	
		\centering
		\includegraphics[width=0.80\linewidth]{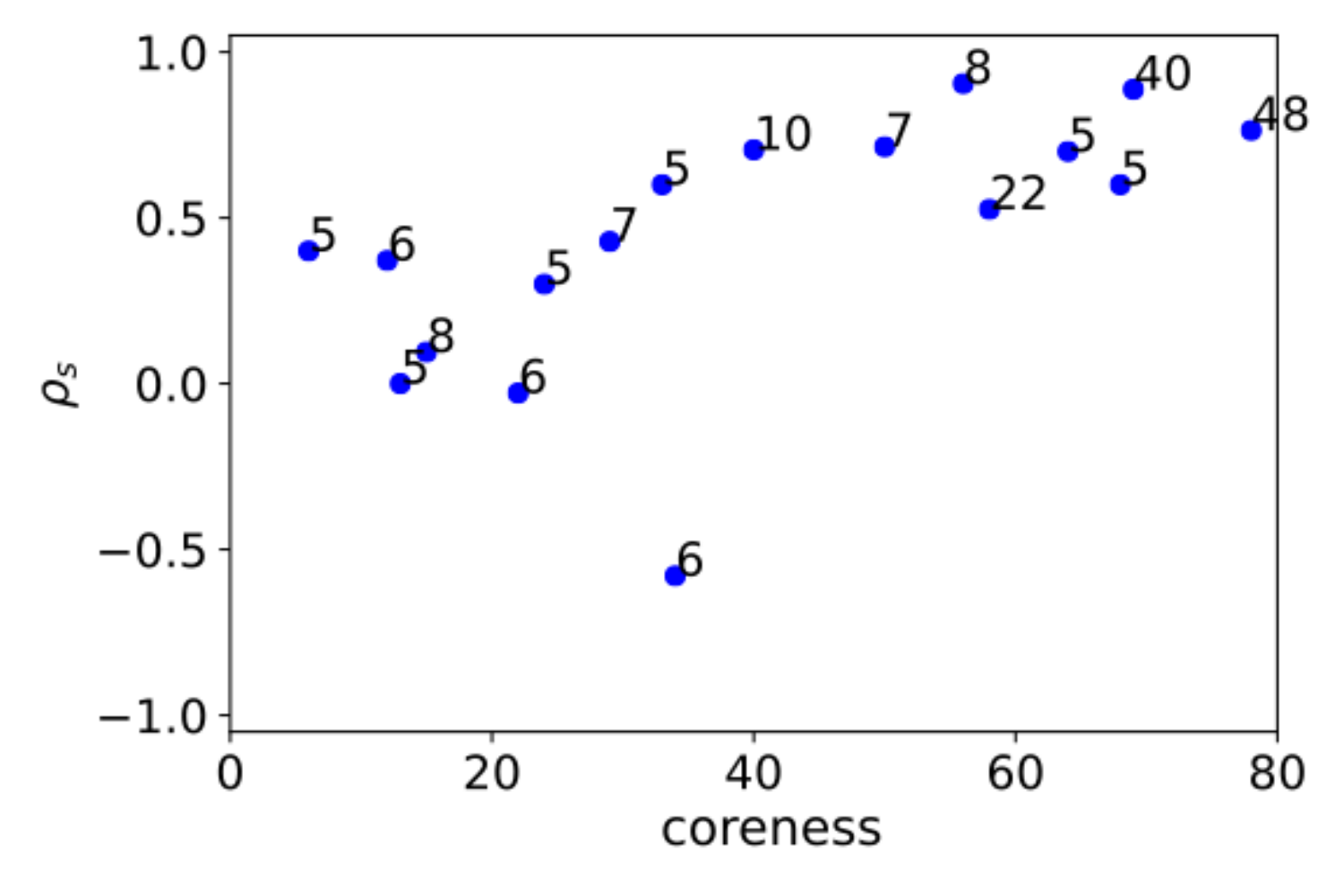}
	\caption[Correlation on core-periphery structure]{Correlation between BHI and PageRank (Spearman correlation coefficient is on the $y$-axis) in different clusters in the same $core$ (having the same coreness value identified by $k-$core decomposition) where coreness is shown in the $x$-axis (from periphery to core left to right).  $k-$core method identifies 2 dense clusters in the core of the mobility network with sizes 40 and 48, corresponding to North American and European cities as shown in Fig. ~\ref{sifig:core_clusters_map} on the map. PageRank becomes less predictive for clusters of cities located in the outer layers of the mobility network. The numbers in each point represent the number of cities in each cluster.}
		\label{sifig:core_clusters_corr}
\end{figure}

\newpage
\clearpage
\section{Exponent $\gamma$ vs correlation coefficients}
\begin{figure}[ht!]
    \centering
    \includegraphics[width=0.7\linewidth]{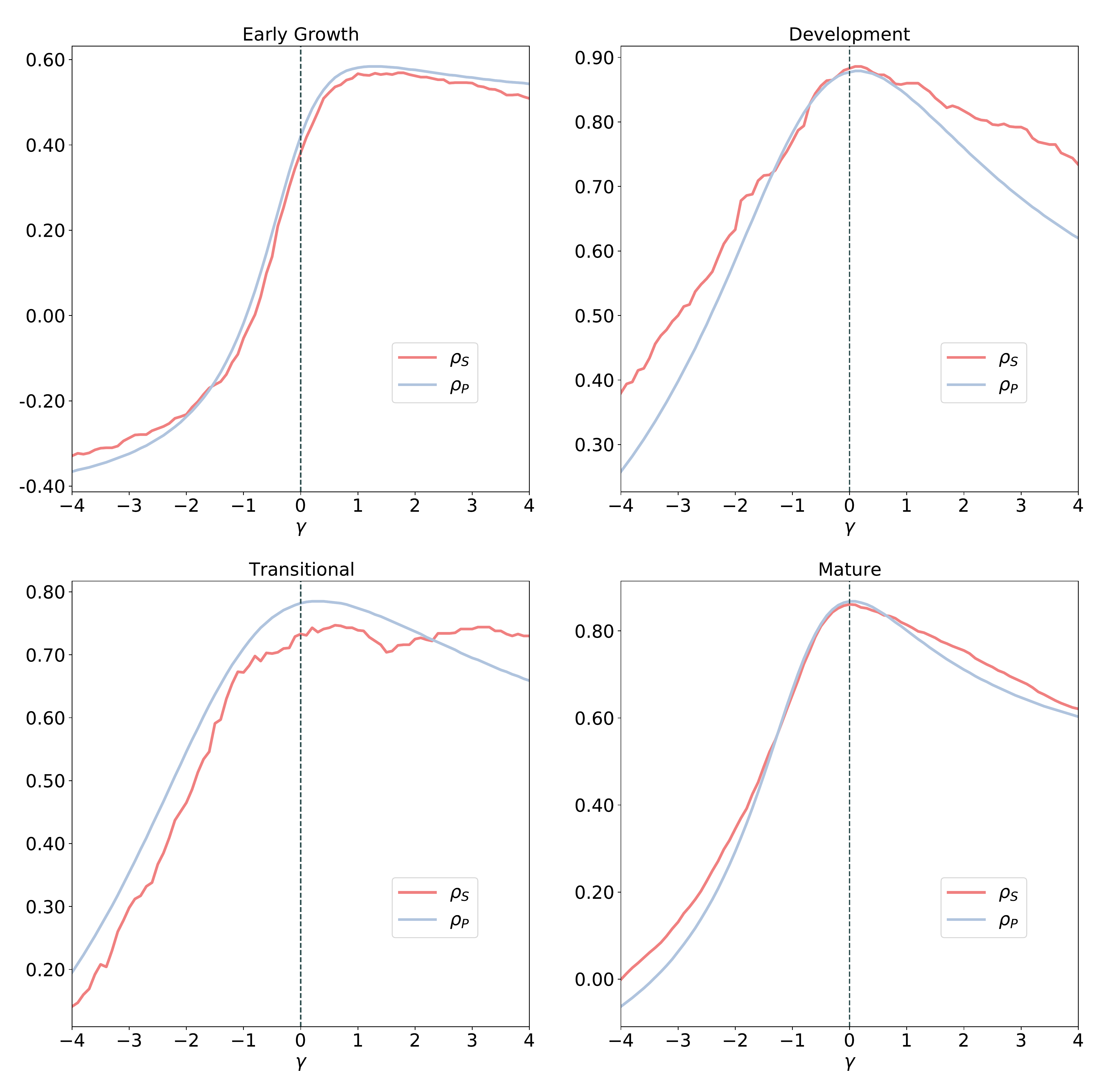}\caption[Dependence of the Spearman's and  Pearson correlation coefficients on the exponent $\gamma$ at Development Stage level]{Dependence of the Spearman's correlation coefficient $\rho_S$ (red) and the Pearson correlation coefficient $\rho_P$ (blue) on the exponent $\gamma$ ($x$-axis) governing how the success of the international connections of one city is relevant to shape its own success. The vertical dashed line represent the correlation existing when just accounting for the centrality of the city, measured by the PageRank, in the mobility network. The different categories correspond to different development stages of the cities.}
    \label{sifig:fit2}
\end{figure}

\begin{figure}[ht!]
	\centering
	\includegraphics[width=0.6\linewidth]{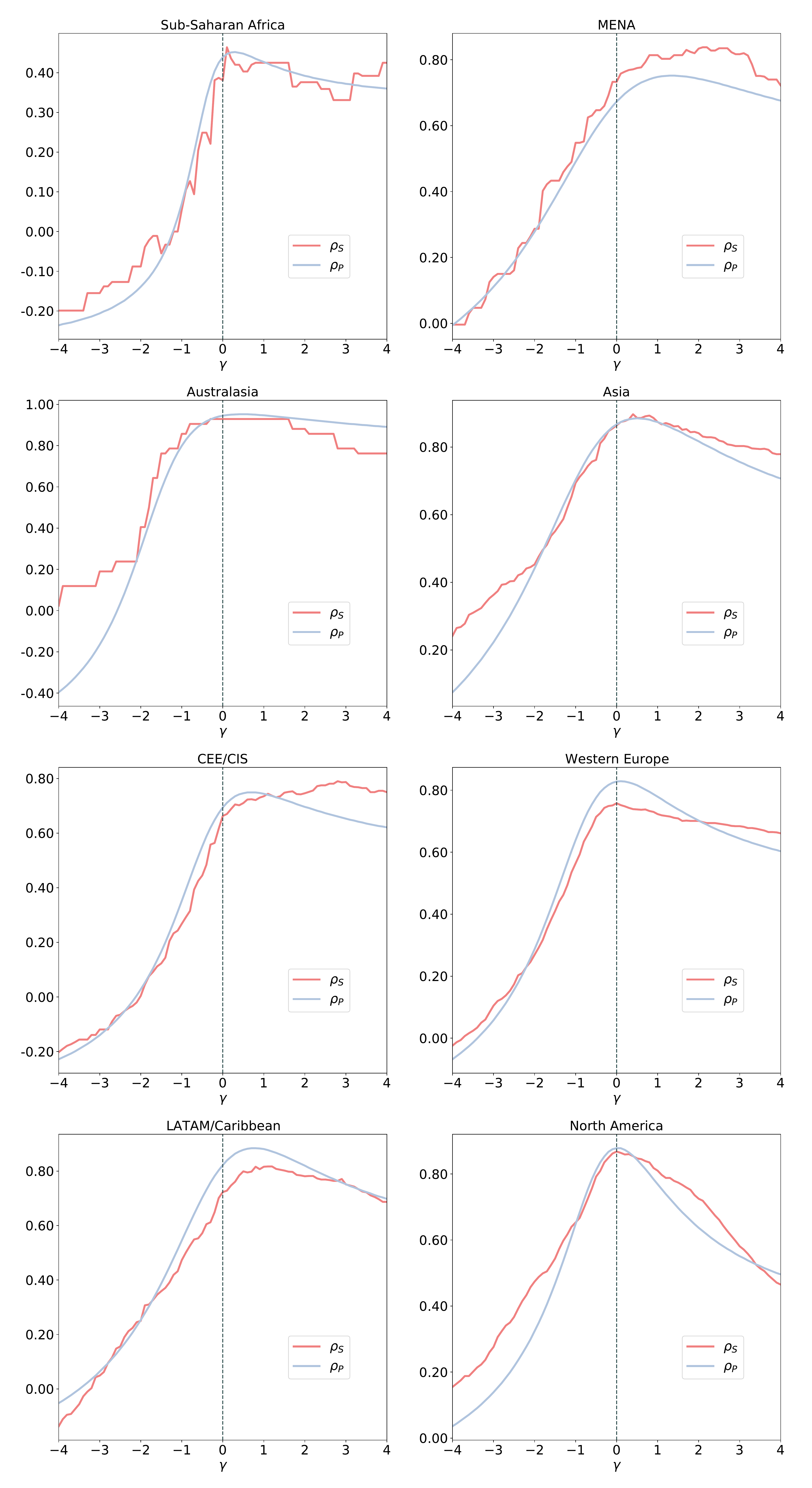}\caption[Dependence of the Spearman's and Pearson's correlation coefficients on the exponent $\gamma$ at Geographical Region level]{Dependence of the Spearman's correlation coefficient $\rho_S$ (red) and the Pearson correlation coefficient $\rho_P$ (blue) on the exponent $\gamma$ ($x$-axis) governing how the success of the international connections of one city is relevant to shape its own success. The vertical dashed line represent the correlation existing when just accounting for the centrality of the city, measured by the PageRank, in the mobility network. The different categories contain cities located in different sub regions across the world.}
	\label{sifig:fit1}
\end{figure}

\begin{figure}[ht!]
	\centering
	\includegraphics[width=0.85\linewidth]{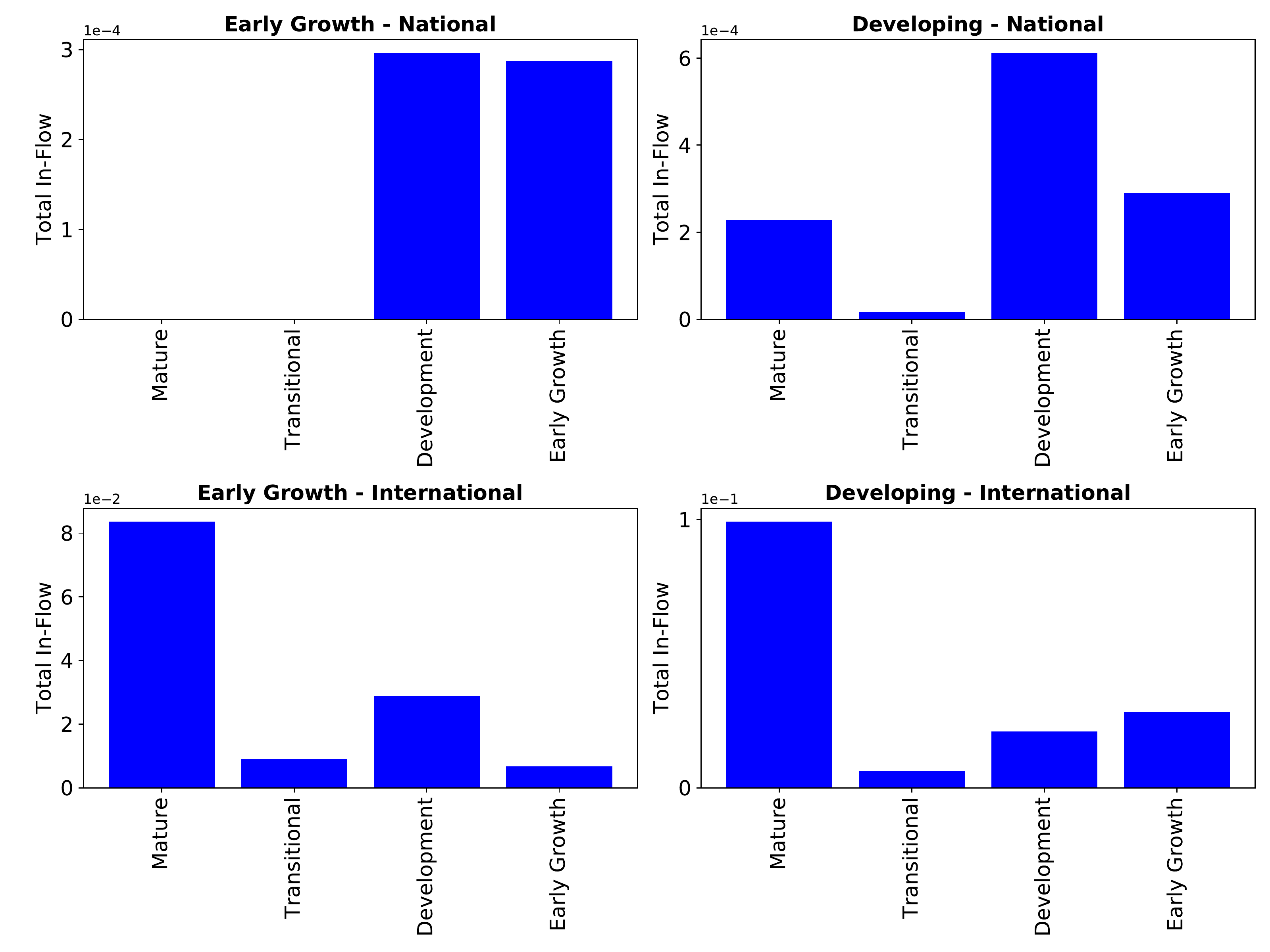}\caption[In-flow distribution of Early Growth and Developing cities]{Breakdown of total in-flow of Early Growth and Developing cities, originating from national (upper panels) and international (lower panels) cities. International flows are dominated by Mature cities, which are insignificant in national inflows.}
	\label{sifig:local_inter_inflow}
\end{figure}

\begin{table}[h]
\scriptsize{
	\begin{filecontents*}{smtable2.csv}
a,b,c,d,e,f,g,h,j,k,l
Development-Stage,Developing,44,1433,0.511,-1.838,0.963,-2.223,0.914,TRUE,0.278
Development-Stage,Early Growth,64,899,0.445,-3.078,0.908,-2.404,0.809,TRUE,0.144
Development-Stage,Mature,131,6395,0.206,-2.529,0.97,-2.705,0.969,TRUE,0.081
Development-Stage,Transitional,29,1032,0.163,-1.473,0.96,-1.365,0.955,TRUE,0.111
Region, North America,64,2903,0.188,-2.556,0.971,-2.739,0.97,TRUE,0.074
Region, LATAM and Caribbean,30,499,0.345,-1.377,0.942,-1.073,0.826,TRUE,0.251
Region, Asia,36,952,0.367,-1.643,0.95,-1.782,0.897,TRUE,0.223
Region, Australasia,8,148,0.44,-1.23,0.941,-0.634,0.719,TRUE,0.358
Region, Western Europe,69,3609,0.393,-2.492,0.778,-2.438,0.71,TRUE,0.158
Region, CEE/CIS,27,903,0.291,-1.847,0.402,-1.672,0.363,TRUE,0.157
Region, MENA,18,630,1.166,-9.011,0.946,-9.735,0.943,TRUE,0.129
Region, Sub-Saharan Africa,16,115,0.296,-1.851,0.999,-1.783,0.998,TRUE,0.16
\end{filecontents*}
\csvreader[tabular={|c
	|c
	|c
	|c
	|c
	|c
	|c|c|c|c|c|},
 table head = \hline \textbf{Category} & \textbf{Classification}&\textbf{$\#$ of cities}&\textbf{$\#$ of flows}
&\textbf{$\hat{\mu}$} 
&\textbf{$\hat{\nu}$} 
&\textbf{$r^2$}
&\textbf{$\nu_0$}
&\textbf{$r^2_0$}
&\textbf{Selected}
&\textbf{$|\hat{\mu} / \hat{\nu}|$} \\\hline,
late after last line=\\\hline]{smtable2.csv}{a=\a,b=\b,c=\c,d=\d,e=\e,f=\f,g=\g,h=\h,j=\j,k=\k,l=\l}%
{\a&\b&\c&\d&\e&\f&\g&\h&\j&\k&\l}
}

\caption[Model Summary]{Results for flow model applied to both maturity and global sub-region city subgroups. The number of cities in each subgroup as well as the number of flows outgoing from these cities that were considered for the correlation analysis are shown. The inferred parameters from Eq. 8 are also shown, along with the corresponding coefficient of determination $r^2$. We additionally display the inferred parameter $\hat\nu_0$ for the restricted model where $\mu=0$, as well as the corresponding coefficient of determination $r_0^2$. The column 'selected' denotes whether or not both the AIC and BIC difference between the two-parameter and one-parameter models are less than zero, indicating that the two-parameter model is preferred by these criteria. Finally, we show the ratio $|\hat\mu/\hat\nu|$, which indicates the relative importance of success versus distance in determining the flows for the corresponding subgroup.}
\label{sitab:modelsummary}
\end{table}